\documentclass[preprint]{revtex4-1}

\usepackage{amsmath}
\usepackage{amssymb}
\usepackage{bm}
\usepackage{color}
\usepackage{hyperref}
\usepackage[normalem]{ulem}
\usepackage[title]{appendix}

\usepackage{titlesec}
\titlespacing\section{0pt}{12pt plus 4pt minus 2pt}{5pt plus 2pt minus 2pt}
\titlespacing\subsection{0pt}{12pt plus 4pt minus 2pt}{2pt plus 2pt minus 2pt}
\titlespacing\subsubsection{0pt}{12pt plus 4pt minus 2pt}{0pt plus 2pt minus 2pt}

\usepackage{graphicx}

\DeclareMathAlphabet{\mathitbf}{OML}{cmm}{b}{it}

\renewcommand{\th}{^{\mbox{\tiny th}}}
\renewcommand{\=}{\!=\!}
\newcommand{\B}[1]{{\bm{#1}}}

\renewcommand{\=}{\!=\!}
\newcommand{\ket}[1]{|#1\rangle}
\newcommand{\bra}[1]{\langle #1|}

\newcommand{\xv}{\mathitbf x}

\newcommand{\calBold}[1]{\mbox{\boldmath${\cal #1}$}}
\newcommand{\mathBold}[1]{\mbox{\boldmath$#1$}}
\newcommand{\dbar}{{\,\mathchar'26\mkern-12mu d}}

\newcommand{\tripleCdot}{\stackrel{\mbox{\bf\scriptsize .}}{:}}

\newcommand{\uv}{\mathitbf u}
\newcommand{\kv}{\mathitbf k}

\begin{document}
\title{Universal disorder-induced broadening of phonon bands:\\ from disordered lattices to glasses}
\author{Eran Bouchbinder${}^{1}$ and Edan Lerner${}^{2}$}
\affiliation{ ${}^1$Chemical and Biological Physics Department, Weizmann Institute of Science, Rehovot 7610001, Israel \\ ${}^2$Institute for Theoretical Physics, University of Amsterdam, Science Park 904, 1098 XH Amsterdam, The Netherlands \vspace{0.6cm}}

\begin{abstract}
\vspace{-1.0cm}
\noindent\rule{16.45cm}{0.4pt}
\vspace{0.2cm}

The translational symmetry of solids, either ordered or disordered, gives rise to the existence of low-frequency phonons. In ordered systems, either crystalline solids or isotropic homogeneous continua, some phonons characterized by different wavevectors are degenerate, i.e.~they share the exact same frequency $\omega$; in finite-size systems, phonons form a discrete set of bands with $n_q(\omega)$-fold degeneracy. Here we focus on understanding how this degeneracy is lifted in the presence of disorder, and its physical implications. Using standard degenerate perturbation theory and simple statistical considerations, we predict the dependence of the disorder-induced frequency width of phonon bands to be $\Delta\omega\!\sim\!\sigma\,\omega\sqrt{n_q}/\sqrt{N}$, where $\sigma$ is the strength of disorder and $N$ is the total number of particles. This theoretical prediction is supported by extensive numerical calculations for disordered lattices --- characterized by topological, mass, stiffness and positional disorder --- and for computer glasses, where disorder is self-generated, thus establishing its universal nature. The predicted scaling of phonon band widths leads to the identification of a crossover frequency $\omega_\dagger\!\sim\!L^{-2/(\dbar+2)}$ in systems of linear size $L$ in $\dbar\!>\!2$ dimensions, where the disorder-induced width of phonon bands becomes comparable to the frequency gap between neighboring bands. Consequently, phonons continuously cover the frequency range $\omega\!>\!\omega_\dagger$, where the notion of discrete phonon bands becomes ill-defined. Two basic applications of the theory are presented; first, we show that the phonon scattering lifetime is proportional to $(\Delta\omega)^{-1}$ for $\omega\!<\!\omega_\dagger$. Second, the theory is applied to the basic physics of glasses, allowing us to determine the range of frequencies in which the recently established universal density of states of non-phononic excitations can be directly probed for different system sizes.\newline
\noindent\rule{16.45cm}{0.4pt}
\end{abstract}

\maketitle

\section{Introduction}
\label{sec:intro}
The low-frequency spectra of condensed matter is generically populated by Goldstone modes that arise due to broken continuous symmetries~\cite{Goldstone_1962,sethna2006statistical}. In~solids, these low-frequency Goldstone modes are long-wavelength phonons (plane-waves) associated with translational symmetry. The distribution of long-wavelength phonons over frequency is of fundamental importance as it determines many thermodynamic and transport properties of solids~\cite{chaikin_lubensky, kittel2005introduction}. In ordered systems, either crystalline solids or isotropic homogeneous continua, some phonons characterized by different wavevectors are degenerate, i.e.~they share the exact same frequency $\omega$. Moreover, in finite-size systems, phonons are quantized into a discrete set of bands with $n_q(\omega)$-fold degeneracy, where $q$ is a phononic band index.

Low-frequency phonons likewise exist in disordered systems, such as disordered crystals/lattices and structural glasses, because they also feature translational symmetry. The presence of disorder, however, is expected to affect the way long-wavelength phonons are distributed over frequency. In particular, the degeneracy of phonons is expected to be lifted by the presence of disorder such that phonon bands broaden, featuring a finite frequency width $\Delta\omega$ each. In this paper we address the following basic question: how do the phonon band widths $\Delta\omega$ depend on the strength of disorder $\sigma$, on the frequency $\omega$, on the original degeneracy level $n_q(\omega)$, and on the system size $N$? To the best of our knowledge, despite its basic nature and the longtime extensive efforts devoted to understanding phonons in disordered systems~\cite{takeno_1971, RevModPhys.44.127, RevModPhys.46.465, Korkusinski_2007,Popescu_2012,Andrew_2017,Sluiter_high_entropy_alloys_2017,PRB_band_unfolding_method_2017}, this question has not been addressed before.

We show that a simple statistical scaling theory predicts $\Delta\omega(\sigma, \omega, n_q, N)\!\sim\!\sigma\,\omega\sqrt{n_q}/\sqrt{N}$. An immediate implication of the existence of finite band widths $\Delta\omega(\sigma, \omega, n_q, N)$ is that there exists a crossover frequency $\omega_\dagger(L)\!\sim\!L^{-\frac{2}{\dbar+2}}$ in a system of linear size $L\!\sim\!N^{1/\dbar}$ in $\dbar\!>\!2$ spatial dimensions (the $\dbar\!=\!2$ case is discussed separately), beyond which the width of phonon bands becomes comparable to the frequency gap between neighboring bands. That is, disordered phonons continuously cover the frequency range $\omega\!>\!\omega_\dagger(L)$, where the notion of discrete phonon bands becomes ill-defined. The theoretical prediction for $\Delta\omega$ is supported by extensive numerical calculations for disordered lattices characterized by topological, mass, stiffness and positional disorder, over a large range of system sizes. Quite remarkably, the analytic prediction for $\Delta\omega$ is shown to be valid also for computer glass-formers, where disorder is self-generated and frustration-induced internal stresses generically emerge, thus establishing its universal nature.

Two basic applications of the developed theory are then presented; first, the theory is applied to phonon dynamics, where the effect of disorder on the scattering lifetime of phonons is considered. In the frequency regime $\omega\!<\!\omega_\dagger(L)$, where phonon bands are well-defined, the theory predicts that the lifetime of phonons becomes finite and is proportional to $(\Delta\omega)^{-1}$. This prediction is quantitatively verified by dynamic phonon scattering calculations in harmonic disordered lattices.

Second, the theory is applied to a basic problem in glass physics; it has been recently shown that structural glasses feature non-phononic low-frequency vibrational modes that are quasilocalized in space and follow a universal density of states $D(\omega)\!\sim\!\omega^4$~\cite{modes_prl, SciPost2016, protocol_prerc, inst_note, all_dimensions_letter, ikeda_pnas, Ikeda_PRE_2018}. Such non-phononic excitations have been hypothesized for decades to be the origin of various universal low-temperature anomalies in glasses~\cite{Pohl_prb_1971, Phillips1972, Anderson1972, RevModPhys.74.991, Klinger_2010_review}. Efforts to observe these glassy excitations of a given frequency $\omega$ have been hampered for a long time because their unique quasilocalized nature is destroyed due to hybridization and mixing with extended phonons that share similar frequencies~\cite{SchoberOligschleger1996,SciPost2016}. Consequently, quasilocalized glassy excitations can be observed in gaps/holes in the phononic spectrum that exist for $\omega\!<\!\omega_\dagger(L)$, explaining the recent observations of~\cite{ikeda_pnas,Ikeda_PRE_2018}. Finally, as $\omega_\dagger(L)\!\to\!0$ in the $L\!\to\!\infty$ limit, quasilocalized glassy excitations cannot be directly observed in the thermodynamic limit.

\section{Scaling theory for disorder-induced broadening of phonon bands}
\label{sec:theory}
The vibrational modes of a solid are determined by the eigenvalue equation ${\calBold M}\ket{\B \psi}\!=\!\omega^2\ket{\B \psi}$, where ${\calBold M} \!\equiv\!\frac{\partial^2{\cal U}}{\partial{\bm x}\partial{\bm x}}$ is the Hessian matrix of dimension $\dbar{N} \times \dbar{N}$ (${\bm x}$ is the vector of $\dbar{N}$ particles' coordinates and ${\cal U}$ is the potential energy), $\ket{\B \psi}$ are the normalized eigenvectors of dimension $\dbar{N}$, and $\omega$ are the vibrational frequencies. Unless noted otherwise, we assume all particles share the same unit mass. There are $\dbar{N}$ eigenvectors, but there can be less vibrational frequencies due to degeneracy. For crystalline (ordered) solids or for isotropic homogeneous continua, the vibrational modes are phonons~\cite{chaikin_lubensky, kittel2005introduction}. Some low-frequency phonons that share the same wavelength/wavenumber are degenerate, i.e.~they also share exactly the same frequency $\omega$; in finite-size systems, phonons form a discrete set of bands with $n_q(\omega)$-fold degeneracy. Our first goal is to understand how this degeneracy is lifted in the presence of disorder.

Let us focus on a single degenerate band of low frequency $\omega$ and denote the eigenvectors within the band by $\ket{\B \psi_i}$, with $i\!=\!1,2,\ldots,n_q$, and the rest (whether degenerate or not) by $\ket{\B \psi_k}$, with $k\!=\! n_q+1,n_q+2,\ldots,\dbar{N}$. We proceed in two steps; first, we use standard degenerate perturbation theory~\cite{footnote3} to obtain an equation for the frequency shifts $\Delta\omega$ due to the presence of disorder characterized by strength $\sigma$. Second, we use the law of large numbers and Wigner's semicircle law for the eigenvalues of random matrices~\cite{Wigner_semicircle} to derive the scaling of $\Delta\omega$ in terms of $\sigma$, $N$, $n_q$, and $\omega$.

We denote by ${\calBold M}^{(0)}$ the Hessian matrix of the system in the absence of disorder. Disorder, which may be realized in many ways (see below), is assumed to be characterized by a typical width $\sigma$, which quantifies its strength. For example, if the disorder is extracted from some distribution, then $\sigma$ is its standard deviation. The presence of disorder modifies the Hessian matrix, which now reads ${\calBold M}\={\calBold M}^{(0)}+\delta{\calBold M}$, where the disorder-induced perturbation $\delta{\calBold M}$ gives rise to shifts $\Delta\omega$ in the originally degenerate frequency $\omega$. That is, the disorder is expected to lift the degeneracy such that each frequency becomes $\omega+\Delta\omega$, with $n_q$ different shifts $\Delta\omega$. The disorder-induced contribution $\delta{\calBold M}$ is a random matrix in which the variability of the elements is determined by $\sigma$.

Within the degenerate band, any linear combination of $\ket{\B \psi_m}$, with coefficients $\beta_m$, is also an eigenvector
\begin{equation}
\label{eq:degenerate_vectors}
{\calBold M}^{(0)}\sum_{m=1}^{n_q} \beta_m \ket{\B \psi_m} = \omega^2 \sum_{m=1}^{n_q} \beta_m \ket{\B \psi_m} \ .
\end{equation}
The linear combination which is relevant for the perturbation problem, which we denote by $\ket{\B \Psi}\!\equiv\!\sum_{m=1}^{n_q} \beta_m \ket{\B \psi_m}$, is not known a priori (there are $n_q$ such linear combinations, but we do not introduce another index for the ease of notation). The crux of standard degenerate perturbation theory is that $\ket{\B \Psi}$, i.e.~the coefficients $\beta_m$, and the frequency shifts $\Delta\omega$ are simultaneously and self-consistently selected according to~\cite{footnote3}
\begin{equation}
\label{eq:degenerate_perturbation_theory}
{\calBold{\tilde M}}\,\ket{\B \beta} = \delta\omega^2\,\ket{\B \beta}  \ .
\end{equation}
Here ${\calBold{\tilde M}}$ is an $n_q \times n_q$ matrix whose elements are given by
\begin{equation}
\label{eq:degenerate_perturbation_theory_1}
{\cal {\tilde M}}_{ij} \equiv \bra{\B \psi_i}\delta{\calBold M}\ket{\B \psi_j} \ ,
\end{equation}
$\ket{\B \beta}$ is an $n_q$-dimensional vector whose components are the coefficients $\beta_m$ and $\delta\omega^2\!\sim\!\omega\Delta\omega$ is the leading order correction to the frequency squared, simply obtained from $(\omega+\Delta\omega)^2$.

The matrix elements ${\cal {\tilde M}}_{ij}$ correspond to contracting the disorder-induced perturbation $\delta{\calBold M}$ of the Hessian matrix with the degenerate normalized modes/eigenvectors $\ket{\B \psi_j}$, from which various scaling properties can be readily derived. First, note that the degenerate phonons $\ket{\B \psi_i}$ are {\em extended} objects, i.e.~their $\dbar{N}$ components are of the same order of magnitude, and are normalized, $\langle{\B \psi_i}\ket{\B \psi_j}\=\delta_{ij}$. This  implies that the components of $\ket{\B \psi_i}$ scale as $1/\sqrt{N}$. Second, ${\cal {\tilde M}}_{ij}$ is a sum of $\dbar{N}$ random numbers characterized by a statistical width $\sigma$, which scales as $\sigma\,\sqrt{N}$ for large $N$, according to the law of large numbers. The last two properties imply that ${\cal {\tilde M}}_{ij}$ scales as $1/\sqrt{N} \times \sigma\,\sqrt{N} \times 1/\sqrt{N} \sim \sigma/\sqrt{N}$. Finally, as $\delta{\calBold M}$ is a second order spatial differential operator contracted with modes of well-defined elastic stiffness $\omega^2$, we also expect ${\cal {\tilde M}}_{ij} \sim \omega^2$. Consequently, we predict that ${\cal {\tilde M}}_{ij}\!\sim\!\sigma\,\omega^2/\sqrt{N}$.

We therefore conclude that the eigenvalue problem defined in Eq.~\eqref{eq:degenerate_perturbation_theory} concerns the eigenvalues of an $n_q \times n_q$ random matrix whose elements scale as $\sigma\,\omega^2/\sqrt{N}$.
The dependence on the degeneracy level $n_q$ is obtained from Wigner's semicircle law which predicts that the eigenvalues of the random matrix ${\calBold{\tilde M}}/\sqrt{n_q}$ follow a semicircle distribution in the large $n_q$ limit~\cite{Wigner_semicircle}. In particular, this distribution has a compact support, which implies that the eigenvalues scale as $\sqrt{n_q}$. Combining this result with ${\cal {\tilde M}}_{ij}\!\sim\!\sigma\,\omega^2/\sqrt{N}$ and $\delta\omega^2\!\sim\!\omega\Delta\omega$, we obtain the final scaling prediction
\begin{equation}
\label{prediction}
\Delta\omega\sim \frac{\sigma\,\omega\sqrt{n_q}}{\sqrt{N}}\ ,
\end{equation}
which is a major result of this paper.

\section{A crossover frequency $\omega_\dagger(L)$}
\label{sec:omega_dagger}

The theory for the band widths $\Delta\omega$ developed above makes sense as long as $\Delta\omega$ does not exceed the gap between neighboring phononic bands. The crossover frequency, denoted hereafter by $\omega_\dagger(L)$, at which the gap between consecutive bands becomes comparable to the band width, is of special interest; it implies that the spectrum of disordered phonons continuously covers the frequency range $\omega\!>\!\omega_\dagger(L)$, with no unoccupied holes. In this frequency range, the notion of phononic bands becomes ill-defined. To calculate the scaling behavior of $\omega_\dagger(L)$ we need to estimate the gaps in frequency between consecutive bands in the ordered system and compare it to $\Delta\omega$. To that aim, we need to express both $\omega_\dagger$ and $\Delta\omega$ in terms of a single quantity that characterizes phononic bands. A natural choice would be simply the serial index $q$ of phononic bands; as will be discussed below, the integer $q$ is directly related to the wavenumber of members of the $q\th$ phononic band for almost --- but not strictly --- all integers.

We start by estimating the degeneracy level $n_q$, which is proportional to the number of solutions to the integer sum of squares problem in $\dbar$-dimensions
\begin{equation}
\label{eq:sum_of_squares}
\sum_{i=1}^\dbar p_i^2 = q\ ,
\end{equation}
where the $p_i$'s and $q$ are integers, and by definition $q\!\ge\!0$. Equation~\eqref{eq:sum_of_squares} is nothing but a relation between the wavenumber squared of a phononic band, which is proportional to $q$, and the squares of its Cartesian components. In dimensions $\dbar\!\ge\!4$, a solution to Eq.~(\ref{eq:sum_of_squares}) exists for any integer $q$, as implied by Lagrange's four-square theorem \cite{Lagrange_four_square}. Importantly, in three dimensions a solution of Eq.~(\ref{eq:sum_of_squares}) exists for most integers $q$ (over 83\% as $q\!\to\!\infty$, where the excluded integers are obtained by Legendre's three-square theorem~\cite{legendre1808essai}), therefore for $\dbar\!\ge\!3$ one can treat $q$ as a band index for scaling purposes (see some comments in this context about $\dbar\=2$ below). For large $q$, $n_q$ is simply given by the surface of a $\dbar$-dimensional sphere in reciprocal space, i.e.~$n_q\!\sim\!q^{\frac{\dbar-2}{2}}$ for $\dbar\!>\!2$. The frequency of phonons belonging to the $q^{\mbox{\tiny th}}$ band follows the dispersion relation of plane-waves $\omega(q)\!\sim\!\sqrt{q}/L$, therefore according to Eq.~(\ref{prediction}) the width of the $q^{\mbox{\tiny th}}$ phonon band reads
\begin{equation}\label{eq:foo00}
\Delta\omega(q, L) \sim \frac{\omega(q) \sqrt{n_q}}{\sqrt{N}} \sim \frac{\displaystyle q^{\frac{\dbar}{4}}}{\displaystyle L^{\frac{\dbar+2}{2}}}\ .
\end{equation}

Next, we estimate the frequency gap $g(q,L)$ between the $q^{\mbox{\tiny th}}$ and the $(q\!+\!1)^{\mbox{\tiny th}}$ bands as a function of $q$. For sufficiently large $q$, $g(q,L)$ is simply given by $d\omega/dq$, i.e.~it follows
\begin{equation}\label{eq:gaps}
g(q, L) \sim \frac{\sqrt{q+1}-\sqrt{q}}{L} \simeq \frac{d\omega}{dq} \simeq \frac{1}{\sqrt{q}L}\ .
\end{equation}
Consequently, the gaps $g(q,L)$ between bands and the band widths $\Delta\omega(q,L)$ become comparable at band index
\begin{equation}
q_\dagger(L) \sim L^{\frac{2\dbar}{\dbar+2}}\,,
\end{equation}
and the crossover frequency $\omega_\dagger(L)$ follows as
\begin{equation}
\label{eq:foo01}
\omega_\dagger(L) \sim \sqrt{q_\dagger}/L \sim L^{-\frac{2}{\dbar+2}}\,.
\end{equation}
We expect this scaling prediction to hold for $\dbar\!>\!2$; all of the concepts discussed above are valid also for $\dbar\=2$, except that a positive integer $q$ is not a proper band index in this case, see additional discussion below (this discussion suggests that Eq.~(\ref{eq:foo01}) is in fact valid also for $\dbar\!=\!2$). Some of the physical implications of this crossover frequency, which evidently vanishes in the thermodynamic limit $L\!\to\!\infty$, will be discussed later in the paper.

\begin{figure}[!ht]
\centering
\includegraphics[width = 0.65\textwidth]{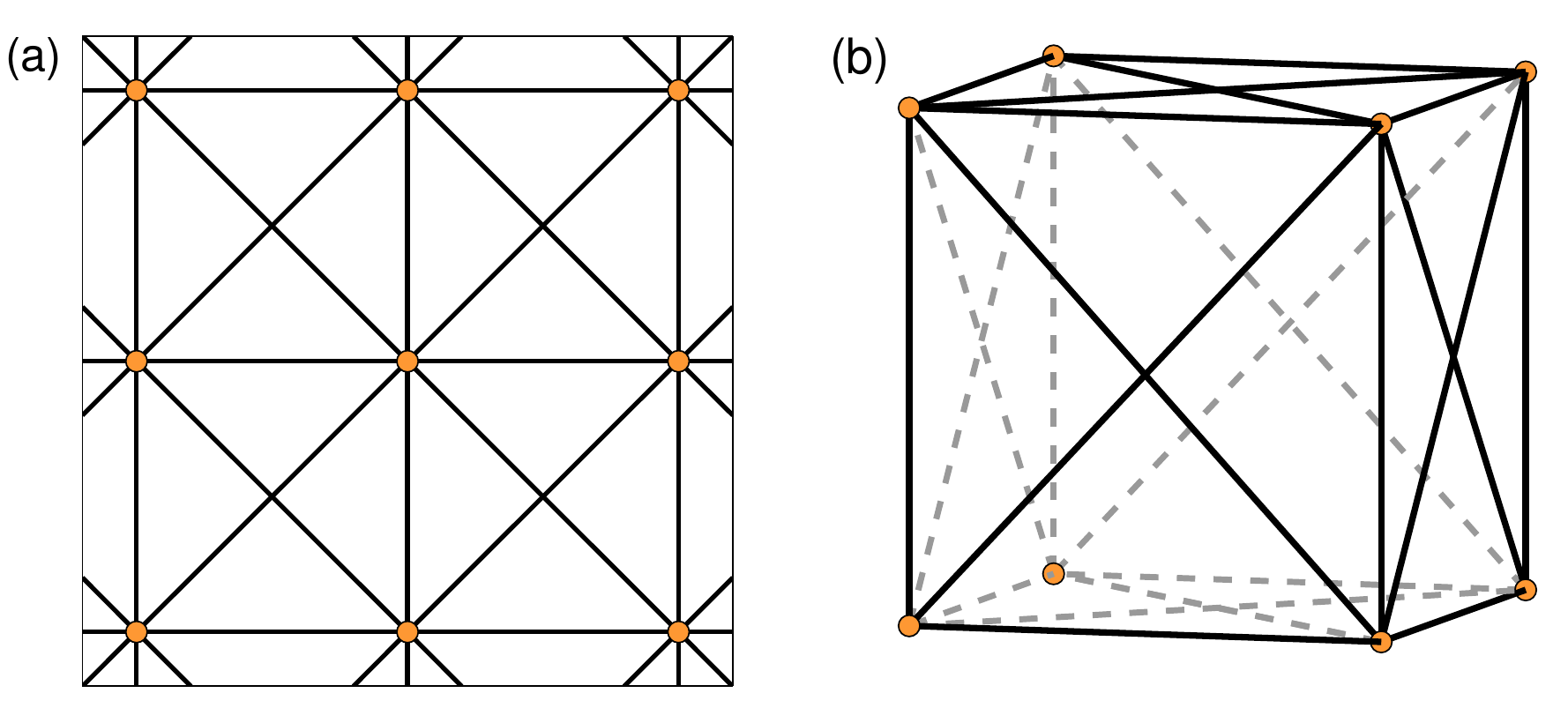}
\caption{\footnotesize The lattices employed in our work: (a) a square lattice in 2D, and (b) a cubic lattice in 3D. Each line connecting two nodes, whether dashed or solid, represents a Hookean spring. Periodic boundary conditions are employed.}
\label{lattices_illustration}
\end{figure}

\section{Numerical experiments}
We next put our theoretical prediction in Eq.~(\ref{prediction}) for the broadening of phonon bands in disordered solids to a direct test. We conducted computer experiments on disordered lattices in which quenched stiffness, mass, and topological disorder are introduced, in addition to measurements performed on generic models of computer glasses, in which disorder is emergent. Details about the models employed, numerical methods and numerical analyses are provided in Appendix~\ref{appendix}.

\begin{figure}[!ht]
\centering
\includegraphics[width = 0.75\textwidth]{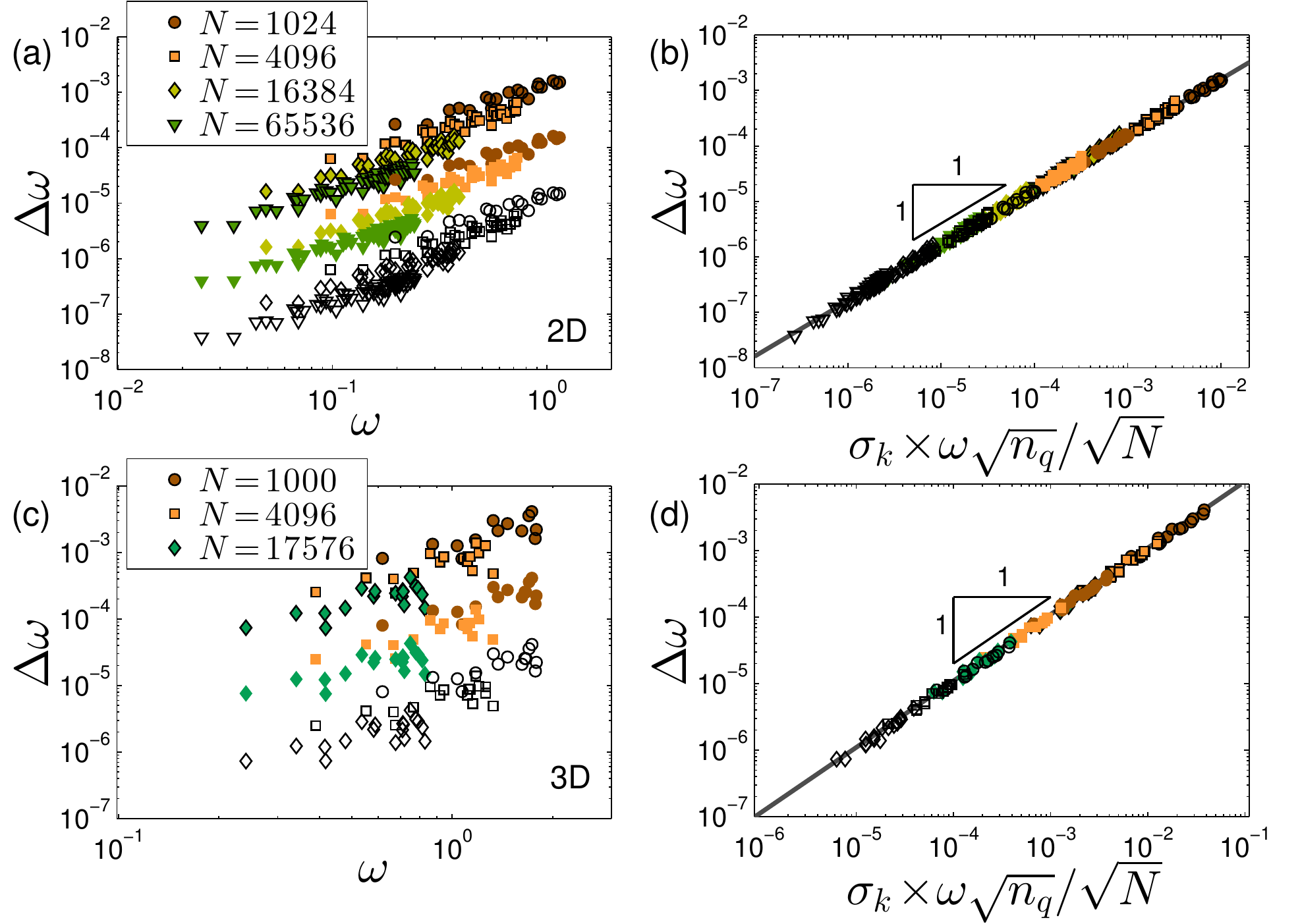}
\caption{\footnotesize Phonon widths measured in (a,b) square and (c,d) cubic lattices of Hookean springs, where random noise of amplitude $\sigma_k$ is introduced in the springs' stiffness. The outlined, full, and empty symbols correspond to $\sigma_k\!=\!10^{-1},10^{-2},$ and $10^{-3}$, respectively.}
\label{stiffness_disorder}
\end{figure}

\begin{figure}[!ht]
\centering
\includegraphics[width = 0.75\textwidth]{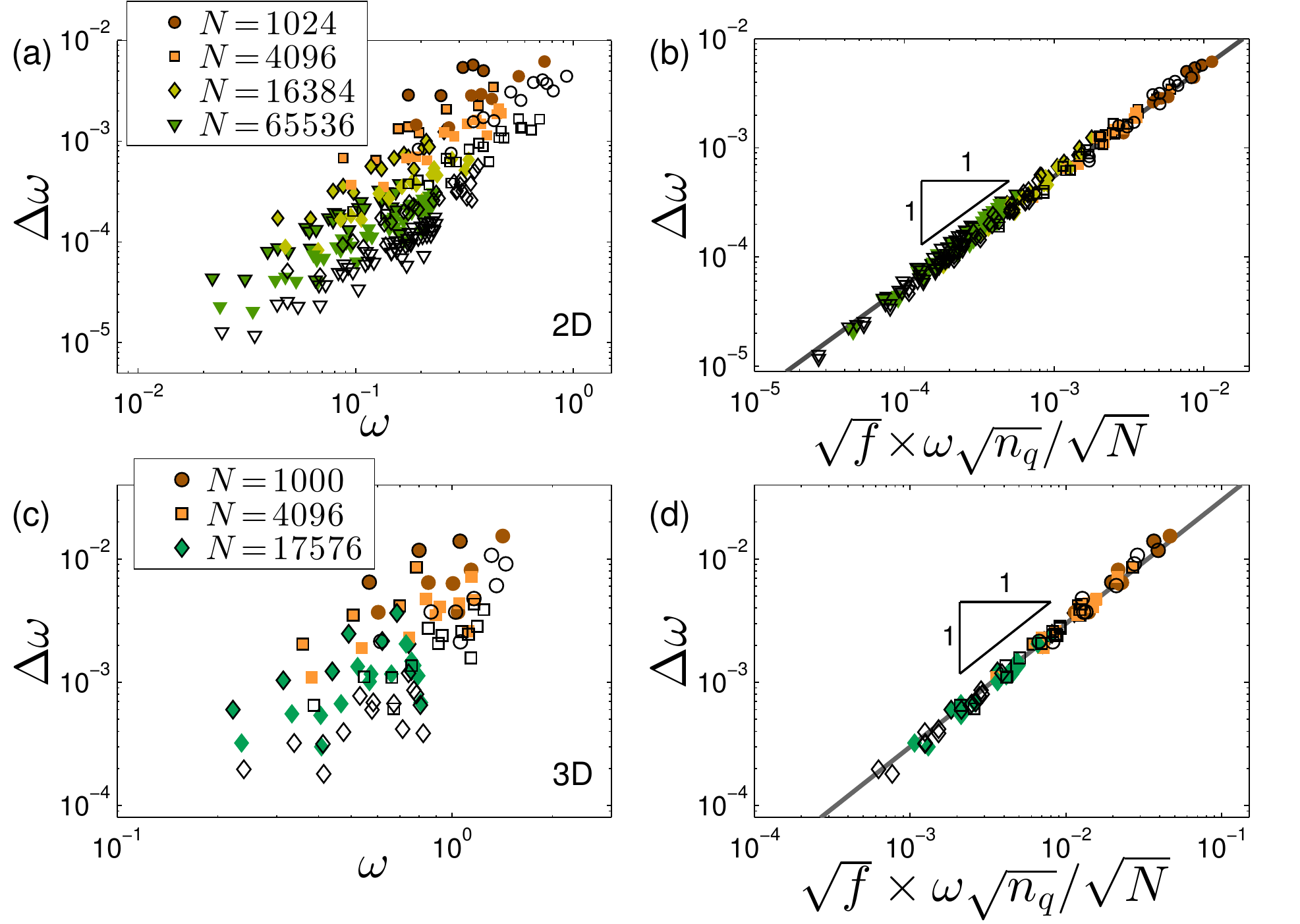}
\caption{\footnotesize Phonon widths measured in (a,b) square and (c,d) cubic lattices of Hookean springs, where a fraction $f$ of the springs are randomly selected and removed. The outlined, full, and empty symbols correspond to $f\!=10\%, 3\%$ and $1\%$, respectively. Note that here $\sqrt{f}$ plays the role of the disorder width, as explained in the text, and hence it is used in the rescaled plots.}
\label{coordination_disorder}
\end{figure}

\subsection{Disordered lattices}
\label{disordered_lattices}
We employ simple lattices of unit masses (unless otherwise stated) connected by relaxed Hookean springs under periodic boundary conditions: square lattices in $\dbar\=2$ (hereafter denoted as 2D), with springs on the diagonals, and cubic lattices in $\dbar\=3$ (hereafter denoted as 3D) with the first (short) diagonals connected by springs, as shown in Fig.~\ref{lattices_illustration}. The connectivities of the square and cubic lattices are $z\!=\!8$ and $z\!=\!18$, respectively. We systematically vary the degree of disorder and the system size, and measure the broadening of phonon bands.

We first assign a random stiffness to each of the lattice springs, drawn from a uniform distribution over the interval $(1\!-\!\sigma_k, 1\!+\!\sigma_k)$. The widths of phonon bands of the resulting stiffness-disordered lattices for $\sigma_k\= 10^{-3}, 10^{-2}$ and $10^{-1}$ are shown in Fig.~\ref{stiffness_disorder}. We find that $\Delta\omega\!\sim\!\sigma_k\,\omega\sqrt{n_q}/\sqrt{N}$, in perfect agreement with the theoretical prediction in Eq.~(\ref{prediction}).

We next randomly select a fraction $f$ of the springs and remove those from the lattice. The widths of phonon bands of the resulting topologically-disordered lattices for $f\!=\!10\%,3\%$ and $1\%$ are shown in Fig.~\ref{coordination_disorder}. Note that in this case $\sqrt{f}$ plays the role of the disorder width because the number of nonzero elements of $\delta{\calBold M}$ is proportional to $f$; this implies that the sum in Eq.~\eqref{eq:degenerate_perturbation_theory_1}, which involves $\sim f\!\times\!\dbar{N}$ random numbers characterized by an $f$-independent statistical width, scales as $\sqrt{f\,N}$ and consequently that ${\cal {\tilde M}}_{ij}$ scales as $1/\sqrt{N} \times \,\sqrt{f\,N} \times 1/\sqrt{N} \sim \sqrt{f}/\sqrt{N}$.

\begin{figure}[!ht]
\centering
\includegraphics[width = 0.75\textwidth]{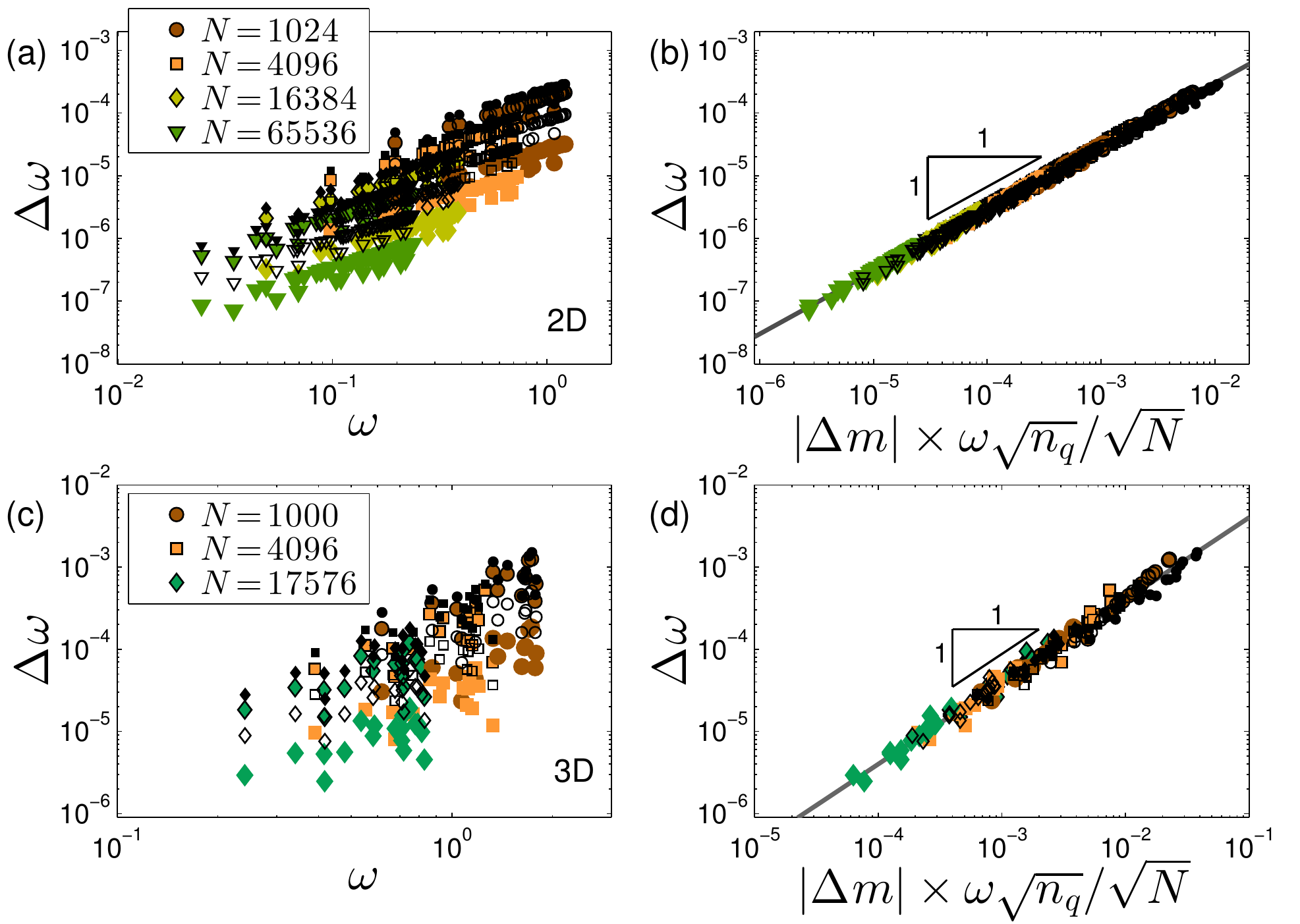}
\caption{\footnotesize Phonon widths measured in (a,b) square and (c,d) cubic lattices of Hookean springs, where 1\% of the particles are assigned a mass that is different by a factor $c_m$ compared to the majority of the particles. The outlined, full, empty and black symbols correspond to $c_m\!=0.94, 0.99, 1.03$ and $1.10$, respectively.}
\label{mass_disorder}
\end{figure}

For our last computer experiment on lattices we randomly selected 1\% of the lattice nodes, and assign them a mass that is different by a factor $c_m$ compared to the mass of the majority of the particles. If masses differ between different degrees of freedom, vibrational frequencies squared are given by eigenvalues of the matrix
\begin{equation}
\calBold{D}_{ij} = \frac{1}{\sqrt{m_im_j}}\calBold{M}_{ij} = \frac{1}{\sqrt{m_im_j}}\frac{\partial^2{\cal U}}{\partial\xv_i\partial\xv_j}\,,
\end{equation}
where $i$ and $j$ are particle indices, $m_i$ and $m_j$ are their associated masses, and $\xv_i$ is the three-dimensional vector of the Cartesian components of $\xv$ pertaining to the $i\th$ particle. We measured the widths of phonon bands for $c_m\!=\!0.94,0.99,1.03$ and 1.10, and the results are displayed in Fig.~\ref{mass_disorder}. We find that the theoretical prediction in Eq.~\eqref{prediction} is exactly followed, with a prefactor given by $|1-c_m|\!\sim\!\Delta m$, where $\Delta m$ is the difference between the minority-species and the majority-species masses.

\subsection{Glass-derived disordered networks}
\begin{figure}[!ht]
\centering
\includegraphics[width = 0.75\textwidth]{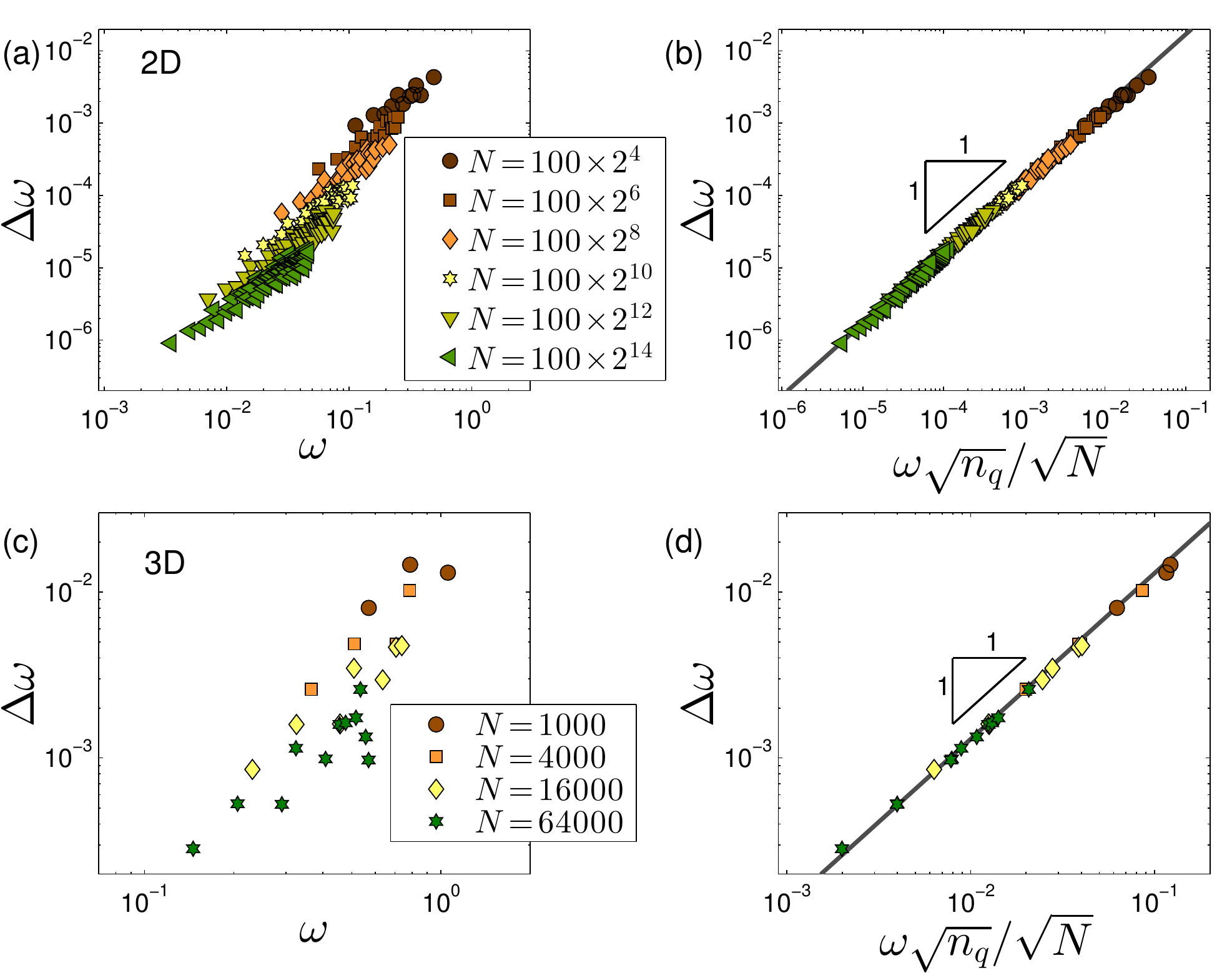}
\caption{\footnotesize Phonon band widths measured in (a,b) 2D and (c,d) 3D glass-derived networks of relaxed Hookean springs with positional disorder, where all springs share the same stiffness, see Appendix~\ref{appendix} for details.}
\label{disordered_networks}
\end{figure}

We next build random networks of Hookean springs by utilizing glassy samples of a generic glass-former, see Appendix~\ref{appendix} for details about the model and the glass preparation protocol. For each glassy sample, we assign a node as the center of each particle, and connect a relaxed Hookean spring with unit stiffness between every pair of particles that interact in the original glass. This procedure leaves us with an ensemble of positionally-disordered networks of Hookean springs, with mean connectivities of $z\!\approx\!6.5$ in 2D, and $z\!\approx\!16.5$ in 3D. We then measured the phonon-band widths in these random networks. The results are plotted in Fig.~\ref{disordered_networks}. We find perfect agreement with the theoretical prediction in Eq.~(\ref{prediction}), despite that these random networks cannot strictly be considered as perturbed lattices.

\subsection{Generic computer glass-formers}
\label{glass_section}

We end this section with presenting results for phonon band widths measured in a generic computer glass model. We employ a 50:50 binary mixture of point-like particles that interact via a pairwise repulsive inverse-power law interaction. This generic model has been shown~\cite{steady_states_with_jacques,protocol_prerc} to reproduce all of the well-known phenomenology of glasses. Glassy samples were prepared by first equilibrating the system in the high temperature liquid phase, and then performing a continuous quench into the glassy phase with a finite quench rate. Further details about the model, analysis and preparation protocol can be found in Appendix~\ref{appendix}.

\begin{figure}[!ht]
\centering
\includegraphics[width = 0.75\textwidth]{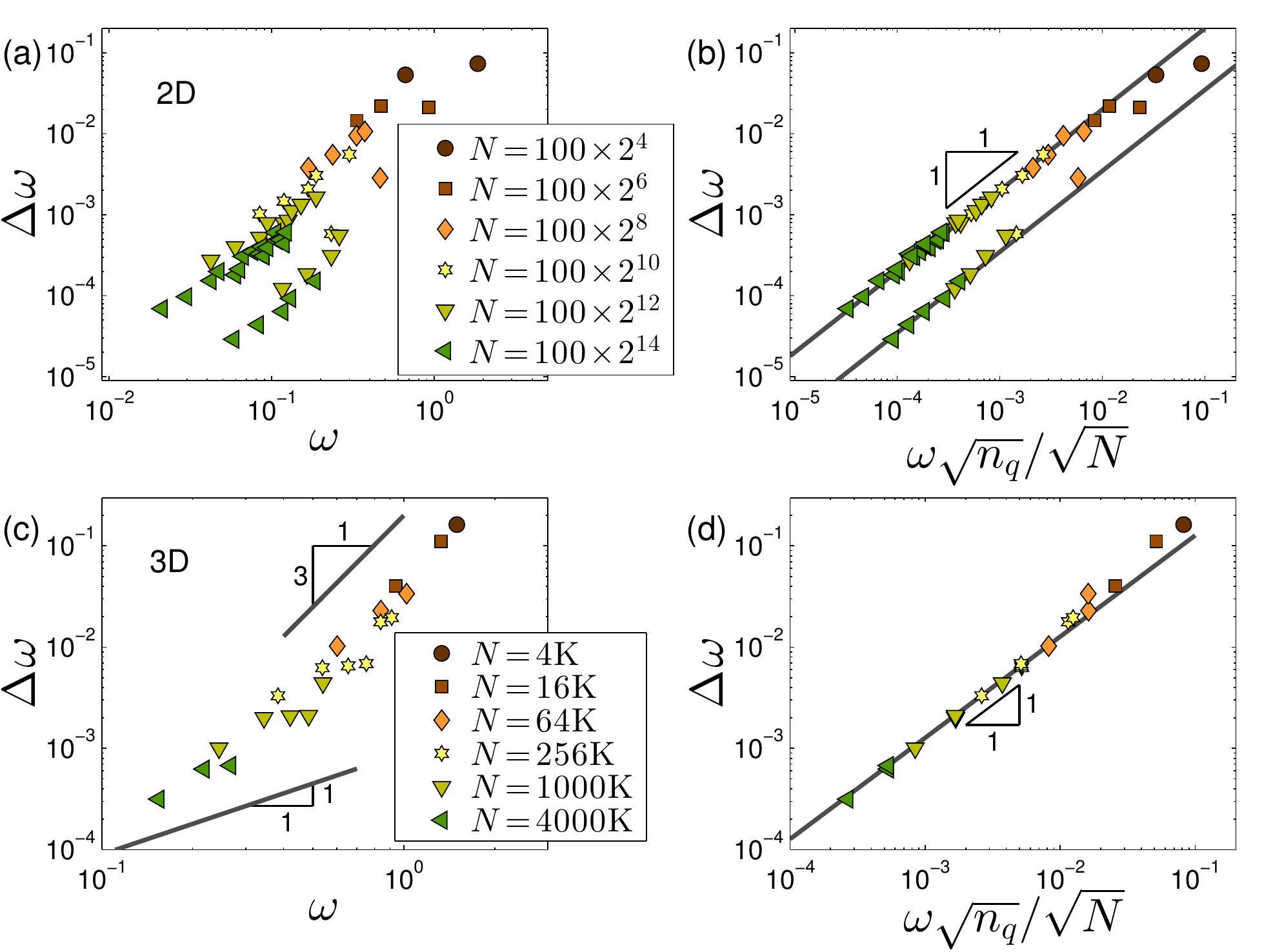}
\caption{\footnotesize Phonon band widths measured in (a,b) 2D and (c,d) 3D model structural glasses, see Appendix~\ref{appendix} for model details. The ratio of the prefactors of the two branches of panel (b), that correspond to shear and sound waves, is found to be $\approx\!5.7$. Note that in panel (c), the raw 3D data of $\Delta\omega$ vs.~$\omega$ appear to follow a power-law. This is a coincidence and the power itself corresponds to a significantly stronger-than-linear relation, as highlighted by the added guides to the eye. When these data are plotted against the rescaled frequency $\omega\sqrt{n_q}/\sqrt{N}$ in panel (d), the predicted linear scaling is demonstrated.}
\label{glass}
\end{figure}

Two key features distinguish this generic structural glass model from the lattice models with quenched disorder discussed above; first, disorder here is spontaneously self-generated during glass formation, rather than being controlled externally. Second, frustration-induced internal stresses generically emerge in glasses~\cite{shlomo}, while they are completely absent (by construction) from the quenched disorder lattice models and the glass-derived disordered networks investigated above. These stresses have been shown to be the origin of several intriguing glassy phenomena, including poor heat transport~\cite{eric_boson_peak_emt}, anomalous scattering of acoustic excitations in the harmonic regime~\cite{schirmacher_prl_2007,Marruzzo2013,Gelin2016}, the existence of quasilocalized glassy modes at the low-frequency end of the vibrational density of states~\cite{modes_prl, SciPost2016, protocol_prerc, inst_note, all_dimensions_letter, ikeda_pnas}, and the singularity of athermal nonlinear elastic moduli~\cite{exist}. It is therefore not a priori clear whether the presence of these internal stresses alters the scaling behavior of the phonon band widths in structural glasses or not.

In Fig.~\ref{glass} we plot the phonon band widths measured for different system sizes in the model structural glasses in 2D (top panels) and 3D (bottom panels). Unlike in the other disordered solids discussed above, here we find two branches when the phonon band widths of the 2D glassy samples are plotted against the rescaled frequency $\omega\sqrt{n_q}/\sqrt{N}$; these correspond to longitudinal (sound) and transverse (shear) phonon bands, where each branch follows the prediction in Eq.~(\ref{prediction}), further establishing its generality and universality. The clear splitting into two branches is not observed in the other models discussed above; we suspect that it is related to the quite different shear $G$ to bulk $K$ elastic moduli ratio, $G/K$, in the different models. In the structural glass model (in 2D) we find $G/K\!\approx\!0.15$, i.e.~$G$ and $K$ are rather well-separated, whereas in the square lattice in 2D $G/K\!=\!2/3$ and in the cubic lattice in 3D $G/K\!=\!3/5$, i.e.~$G$ and $K$ are of similar magnitude. These differences, while suggestive, do not entirely explain why in the disordered lattices case the prefactors in the $\Delta\omega$ scaling relations for the longitudinal (sound) and transverse (shear) phonon bands are essentially indistinguishable.

The 3D data plotted in Fig.~\ref{glass}c,d correspond only to the scaling of phonon band widths of transverse (shear) phonons. This is the case for two reasons: first, for small system sizes it is impossible to reliably extract the width of longitudinal (sound) phonon bands since different bands are not well-separated from each other. For larger systems, we cannot access any longitudinal (sound) phonon bands since those occur at frequencies that are too high for us to compute due to computational bottlenecks. Nevertheless, the scaling we observe for the transverse (shear) phonon band widths in 3D structural glasses convincingly follows the theoretical prediction as well. With this we conclude our numerical experiments, which verified the universal validity of the theoretical prediction in Eq.~\eqref{prediction}. We next apply the theory to two basic problems in disordered systems.

\section{Application to phonon dynamics: Phonon scattering lifetime}

Having established that the disorder-induced broadening of phonon band frequency widths follows the theoretical prediction in Eq.~\eqref{prediction}, independently of the type of disorder, we next discuss its implications for phonon dynamics. In ordered systems, phonons propagate indefinitely in the harmonic regime, i.e.~in the absence of anharmonicity. The presence of disorder is expected to qualitatively change this picture, i.e.~we expect pure phonons that are eigenmodes of the original ordered system to feature a finite lifetime even in the harmonic regime. In the frequency range $\omega\!<\!\omega_\dagger(L)$, where phononic bands are well-defined, we expect an excited phononic state that belongs to the original degenerate band --- which is {\em not} an eigenmode of the disordered system --- to have significant projections only on members of its own band --- which is not degenerate anymore in the presence of disorder. As the spectral width of the band in the presence of disorder is $\Delta\omega$, we expect the lifetime to be proportional to $(\Delta\omega)^{-1}$.

To test this prediction, we perform numerical scattering simulations on harmonic square lattices with quenched stiffness disorder of strength $\sigma_k$, as described in Sect.~\ref{disordered_lattices}. At time $t\=0$, we introduce at the lattice positions ${\bm r}_i$ ($i$ is the particle index) velocity vectors of the form
\begin{equation}
\dot{\uv}_i(t\=0) = {\bm a}\,\sin(\kv\!\cdot\!{\bm r}_i )\,,
\end{equation}
which correspond to pure modes of the perfectly ordered lattice. Here the wavevector is given by $\kv\!=\!(2\pi/L)(p_x {\bm e}_x + p_y {\bm e}_y)$, where $p_x,p_y$ are integers such that $p_x^2+p_y^2\=q$, cf.~Eq.~\eqref{eq:sum_of_squares}, and ${\bm e}_x, {\bm e}_y$ are the corresponding unit vectors. The normalized polarization vector ${\bm a}$ is defined by ${\bm a}\!\cdot\!\kv\!=\!0$ and ${\bm a}\!\cdot\!{\bm a}\!=\!1$, i.e.~we consider shear plane-waves, which for small $p_x, p_y$ correspond to the lowest frequency excitations.

\begin{figure}[!ht]
\centering
\includegraphics[width = 0.75\textwidth]{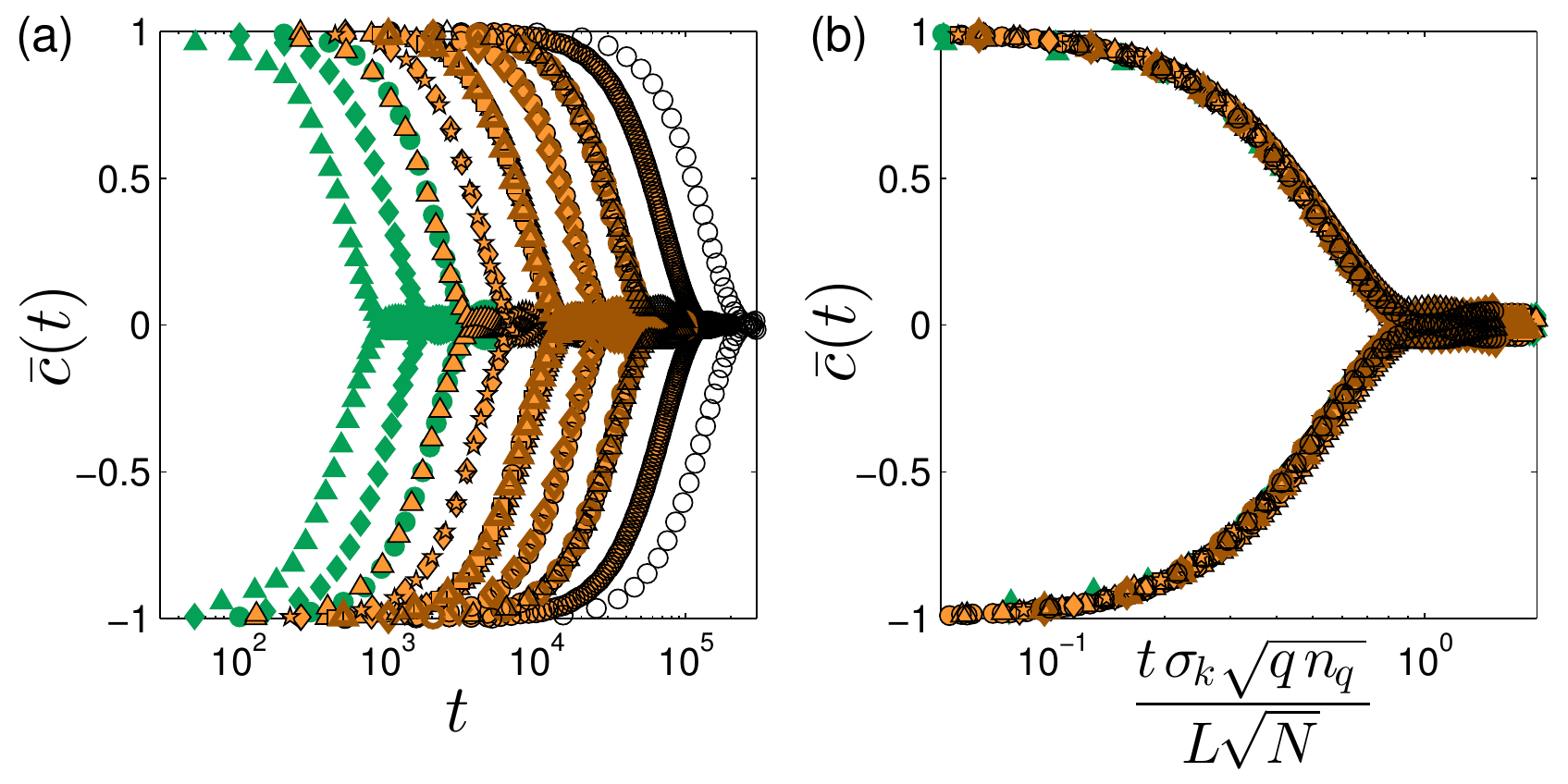}
\caption{\footnotesize (a) Mean envelopes of the velocity autocorrelation function of phonons, $\bar{c}(t)$, calculated in harmonic square lattices with uniformly distributed quenched stiffness disorder of strength $\sigma_k$, cf.~Sect.~\ref{disordered_lattices}. Full, outlined, bold, and empty symbols correspond to lattices of $N\!=\!1600,6400,25600$ and $102400$ nodes, with band indices $q\!=\!1,2,4,8$ and $16$ represented by circles, squares, diamonds, stars, and triangles, respectively. The strength of disorder $\sigma_k$ is varied between $0.05$ and $4.0$, in steps of factor of two, for $N\!=\!6400$. For each set of parameters, averages were taken over 1000 independent disorder realizations. (b) Same as in (a), but plotted against the rescaled time $t\, \sigma_k\sqrt{q\,n_q}/(L\sqrt{N})\!\sim\!t\,\Delta\omega$, demonstrating an essentially perfect data collapse.}
\label{scattering}
\end{figure}

The $2N$-dimensional velocity vector $\dot{\uv}(t)$, composed of the Cartesian velocity vectors $\{\dot{\uv}_i(t)\}$ of all of the particles, is calculated for $t\!>\!0$ using harmonic lattice dynamics and the scattering lifetime is probed through the time-evolution of the velocity auto-correlation function
\begin{equation}
c(t)=\frac{\dot{\uv}(t)\!\cdot\!\dot{\uv}(t\=0)}{\dot{\uv}(t\=0)\!\cdot\!\dot{\uv}(t\=0)} \,,
\label{eq:autocorrelation}
\end{equation}
The mean envelopes (i.e.~the amplitude of the function, disregarding its oscillatory part) of $c(t)$, $\bar{c}(t)$, are plotted for various system sizes, wavenumbers, disorder strength and degeneracy levels in Fig.~\ref{scattering}a, see caption for details. The theoretical prediction is that the inverse scattering lifetime, i.e.~the so-called damping coefficient $\Gamma$, scales with $\Delta\omega$. Consequently, we expect the velocity auto-correlation function $c(t)$ to collapse onto a master curve once plotted against $t\,\Delta\omega$. This is explicitly demonstrated in Fig.~\ref{scattering}b, where $\bar{c}(t)$ is plotted against $t\,\sigma_k\sqrt{q\,n_q}/(L\sqrt{N})\!\propto\! t\,\sigma_k\,\omega\sqrt{n_q}/\sqrt{N}\!\sim\!t\,\Delta\omega$, where the phononic dispersion relation $\omega\!\propto\!\sqrt{q}/L$ was used. This result establishes that the phonon scattering lifetime for $\omega\!<\!\omega_\dagger(L)$ is indeed determined by $(\Delta\omega)^{-1}$.

Applying these results to 3D, the damping coefficient is predicted to scale as $\Gamma(\omega)\!\sim\!\Delta\omega\!\propto\!\omega\sqrt{n_q(\omega)}\!\propto\!\omega^{3/2}$. The scaling relation $\Gamma(\omega)\!\sim\!\omega^{3/2}$ is manifestly different from the conventional Rayleigh scattering prediction $\Gamma(\omega)\!\sim\!\omega^{4}$ (and its recently discussed generalization for glasses~\cite{Gelin2016}). The resolution of this apparent discrepancy is that the $\Gamma(\omega)\!\sim\!\omega^{3/2}$ scaling dominates the frequency range $\omega\!<\!\omega_\dagger(L)$, while we expect Rayleigh scattering scaling $\Gamma(\omega)\!\sim\!\omega^{4}$ to be observed in the frequency range $\omega\!>\!\omega_\dagger(L)$ (at least for disordered lattices). Consequently, the frequency scale $\omega_\dagger(L)$ is predicted to control the finite-size-dependent crossover between the two scaling regimes. These predictions will be systematically addressed elsewhere.

\section{Application to glass physics: Coexistence of quasilocalized glassy modes and phonons}

The theoretical prediction in Eq.~\eqref{prediction} for the disorder-induced broadening of phonon band frequency widths has important consequences not only for phonon dynamics, but also for non-phononic vibrational modes in structural glasses, which are of fundamental importance in the physics of glasses. It has been recently discovered~\cite{modes_prl} that non-phononic vibrational modes populate the low-frequency tails of the vibrational density of states of structural glasses, whose frequencies are distributed according to a universal gapless $\omega^4$ law, independently of microscopic details. It is now broadly accepted that these soft non-phononic excitations are spatially quasilocalized~\cite{modes_prl, SciPost2016, all_dimensions_letter}, i.e.~they feature a disordered core characterized by a localization length of a few atomic sizes in linear dimension and are accompanied by either a power-law decay (unlike Anderson-localized modes that appear at the high-frequency end of the phononic spectrum and decay exponentially in space~\cite{anderson_localization}), or an extended, wave-like background that spans the entire system \cite{Schober_jop_2004,Schober_Ruocco_2004,ikeda_pnas}. Furthermore, these soft non-phononic excitations are known to emerge from the presence of frustration-induced internal stresses in the glass~\cite{inst_note}. The gapless $\omega^4$ law of these soft quasilocalized glassy modes (QLGMs) had been predicted decades ago~\cite{soft_potential_model_1987,soft_potential_model_1991}, but only recently has it been firmly established for the first time in simple, finite-size computer glass-formers.

Notwithstanding, one of the central questions left unresolved concerns the survival of QLGMs as a function of the system size $L$ and in particular in the thermodynamic limit $L\!\to\!\infty$. To address this important question one needs to understand how phonons cover the frequency axis as a function of $L$ and whether QLGMs retain their quasilocalized nature when they share the {\em same} frequencies with phonons. The former was completely addressed in Sect.~\ref{sec:theory}, where we have derived the system-size scaling of the crossover frequency $\omega_\dagger\!\sim\! L^{-2/5}$ in 3D, above which phonon bands overlap and merge, leaving the density of vibrational modes free of gaps and holes. In particular, in the thermodynamic limit the crossover frequency vanishes $\omega_\dagger\!\to\!0$, implying that phonons then fully occupy the entire low-frequency regime. Consequently, the possibility that QLGMs and phonons coexist in the thermodynamic limit in 3D crucially depends on whether QLGMs can share the {\em same} frequencies with phonons, without undergoing hybridizations and mixing that destroy their quasilocalized nature. This point is addressed next.
\begin{figure}[!ht]
\centering
\includegraphics[width = 0.77\textwidth]{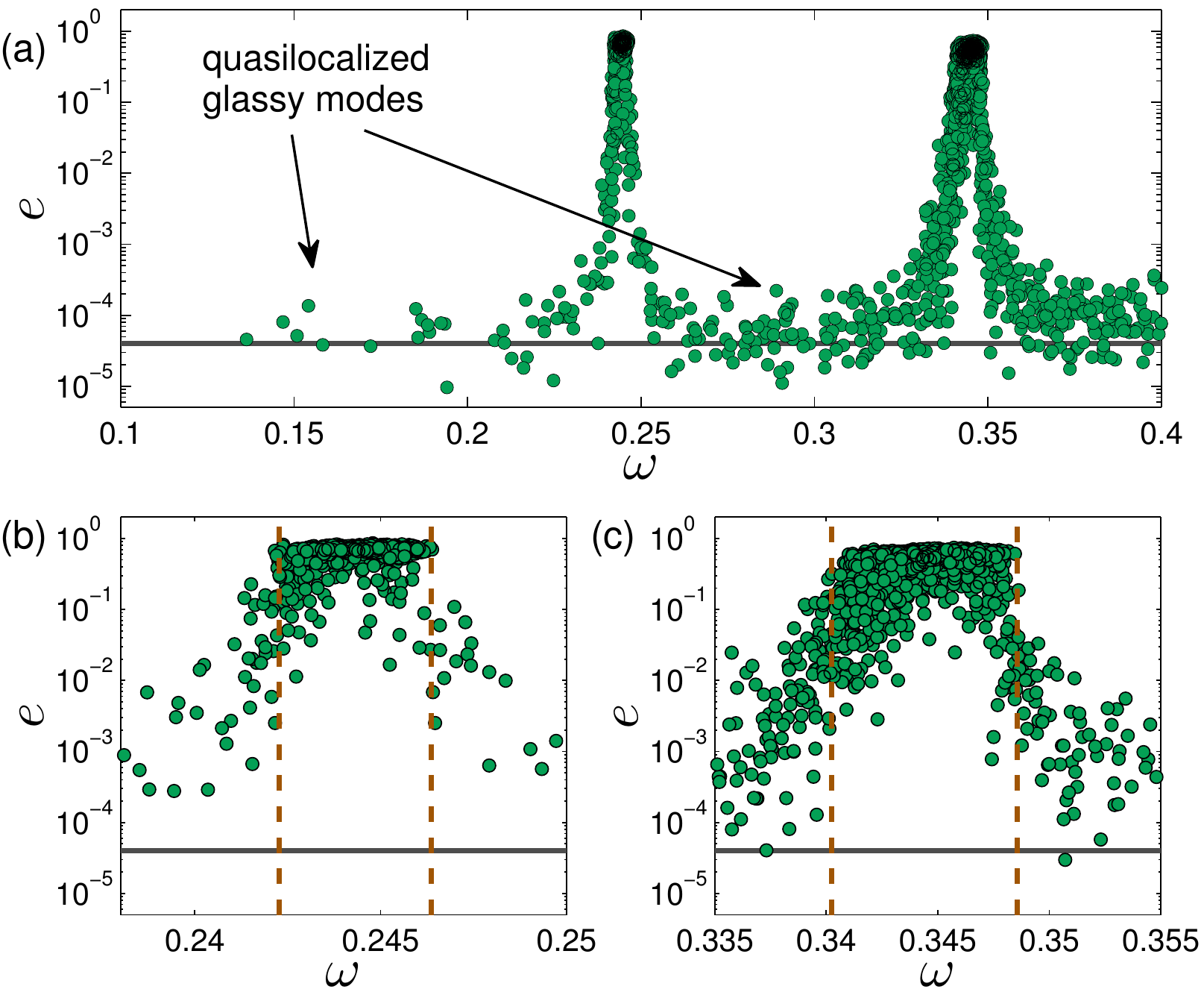}
\caption{\footnotesize Participation ratio $e$ (see text for definition) of normal modes of frequency $\omega$, measured in our 3D computer glass of $N\!=\!10^6$ particles. Panel (a) shows the entire calculated frequency range, whereas panels (b) and (c) are zoomed in on the first and second phonon bands, respectively. The horizontal line indicates the participation ratio of QLGMs, and the dashed vertical lines engulf a frequency window of 4 standard deviations of each phonon band.}
\label{hybridization_fig}
\end{figure}

A systematic, large-scale study of hybridization and mixing phenomena of QLGMs and phonons, including its frequency and system-size dependence, is left for a future report; here we show preliminary data indicating that strong hybridizations of QLGMs with phonons of similar frequencies occur, and, consequently, no QLGMs survive within phonon bands. We generated $100$ independent glassy samples of $N\=10^6$ particles in 3D, and calculated the first $40$ vibrational modes $\mathBold{\Psi}$ with nonzero vibrational frequencies, see Appendix~\ref{appendix} for details about the model and the calculation. The degree of localization of the calculated vibrational modes is effectively captured by the participation ratio $e$, defined as
\begin{equation}\label{participation_ratio_definition}
e \equiv \frac{1}{N\sum_i ({\mathBold \Psi}_i\cdot{\mathBold \Psi}_i)^2}\,,
\end{equation}
where ${\mathBold \Psi}_i$ is the three-dimensional vector of the Cartesian components of ${\mathBold \Psi}$ pertaining to the $i^{\mbox{\tiny th}}$ particle. Extended vibrational modes, such as phonons, are characterized by $e\!\sim\!{\cal O}(1)$. Localized or quasilocalized modes, such as QLGMs, are characterized by $e\!\sim\!{\cal O}(N^{-1})$. In Fig.~\ref{hybridization_fig} we scatter-plot the participation ratio $e$ of the vibrational modes vs.~their frequency $\omega$; panel (a) shows the entire calculated frequency range, whereas panels (b) and (c) zoom into the first and second phonon bands, respectively. The horizontal continuous lines mark the characteristic value of the participation ratio of QLGMs, shown in~\cite{modes_prl, SciPost2016} to scale as $N^{-1}$. The dashed vertical lines in panels (b) and (c) enclose a frequency window of 4 standard deviations of each phonon band, obtained from the density of vibrational modes as described in Sect.~\ref{glass_section} above.

The presented data indicate that within the frequency intervals covered by phonon bands, QLGMs have a very strong tendency to hybridize with phonons of those bands, which destroys their quasilocalized nature. In the presented ensemble of $100$ glassy samples of $N\!=\!10^6$, we expect (see Appendix~\ref{nonlinear_modes_appendix} for details about this estimation) that in the absence of phonons we would find $\approx\!3.3$ QLGMs within the frequency interval covered by the first phonon band, and $\approx\!27$ QLGMs within the frequency interval covered by the second phonon band. However, as clearly seen in Figs.~\ref{hybridization_fig}b,c, we do not find a {\em single} QLGM within the frequency intervals covered by both the first and second phonon bands.

The strong hybridizations of QLGMs seen within phonon bands imply that they can only exist at frequencies $\omega\!\lesssim\!\omega_\dagger$, i.e.~below the crossover frequency, where gaps in the spectrum open between subsequent phonon bands. Can QLGMs exist at such low frequencies? The answer to this question was put forward in~\cite{modes_prl}, where it was shown that the characteristic scale $\omega_g$ at which QLGMs start to appear depends on the linear system size $L$ as $\omega_g\!\sim\! L^{-3/5}\!\lesssim\!\omega_\dagger$ in 3D, and generally as $L^{-\dbar/5}$ in $\dbar$ dimensions. This scaling is a direct consequence of the universal distribution of QLGMs that varies as $\omega^4$ independently of spatial dimension \cite{all_dimensions_letter}, and of the quasilocalized nature of those modes. Combining the scaling results for $\omega_\dagger$ and $\omega_g$, we conclude that \emph{QLGMs and phonons coexist} for frequencies $\omega$ in the interval
\begin{equation}
\omega_g \sim L^{-3/5} \lesssim \omega \lesssim L^{-2/5} \sim \omega_\dagger \quad \mbox{in 3D} \ .
\end{equation}

Our results suggest that in the thermodynamic limit $L\!\to\!\infty$ in 3D all QLGMs cease to exist as {\em harmonic} vibrational modes, at odds with the recent claims of~\cite{ikeda_pnas,Ikeda_PRE_2018}, that argue that these two types of low-frequency modes can be distinguished by virtue of their participation ratio in the thermodynamic limit. While quasilocalized excitations certainly do not disappear due to the presence of phonons with similar frequencies~\cite{luka, SciPost2016, manning_defects} --- as they correspond to particularly soft structures within the glass~\cite{thermal_energies} ---, their realization as harmonic vibrational modes (i.e.~as eigenvectors of the Hessian matrix) with participation ratios of order $N^{-1}$ becomes impossible due to strong hybridizations with phonons. Consequently, their detection in the thermodynamic limit by conventional techniques is hindered. The important result that {\em harmonic} QLGMs do not coexist with phonons in the thermodynamic limit is further strengthened by the existence of {\em nonlinear} QLGMs within the frequencies occupied by phonon bands, which is demonstrated in Appendix~\ref{nonlinear_modes_appendix} using the framework introduced in~\cite{SciPost2016}.

In dimensions $\dbar\!>\!3$, the coexistence window of QLGMs and phonons follows $L^{-\dbar/5}\!\lesssim\!\omega\!\lesssim\! L^{-2/(2+\dbar)}$, i.e.~it grows in relative terms with increasing dimension. We note, however, that for dimensions $\dbar\!>\!5$ this coexistence window extends well-below the longest-wavelength phonons, of frequencies $L^{-1}$. This means that for $\dbar\!>\!5$ QLGMs appear undisturbed by phonons between $L^{-\dbar/5}\!\lesssim\!\omega\!\lesssim\! L^{-1}$. Finally, we comment on the behavior expected in 2D structural glasses; under assumptions spelled out and motivated in Appendix~\ref{sec:2D}, we expect $\omega_\dagger\!\sim\! L^{-1/2}$ in 2D. On the other hand, the onset of glassy modes is expected to follow $L^{-2/5}$ in 2D (with possible logarithmic corrections, see discussion in~\cite{cge_paper}), i.e.~it is larger than the crossover frequency $\omega_\dagger$. This means that in 2D, above some system size, we expect harmonic QLGMs to be unobservable altogether, and no coexistence regime with phonons to exist, as indeed reported in~\cite{ikeda_pnas}.

\section{Summary and discussion}
\label{sec:summary}

In this work we have theoretically derived the dependence of the disorder-induced frequency widths of phonon bands $\Delta\omega$ on the strength of disorder, on the number of phonons in a band, on the band's frequency, and on the number of particles in solids. Our result, which is obtained using degenerate perturbation theory and simple statistical considerations, was then validated against computer experiments on simple lattices with quenched stiffness, mass, and topological disorder, and on generic computer models of structural glasses, both in two and three dimensions, and for a wide variety of system sizes. In all cases we found excellent agreement between our theoretical prediction and the numerical data, establishing the universal nature of the theory. Of particular interest is the robustness of our scaling theory to the presence of self-generated disorder and frustration-induced internal stresses in structural glasses, which are known to affect many phonon-related glassy phenomena~\cite{eric_boson_peak_emt,Gelin2016}.

The derived broadening of phonon band widths with increasing frequency gives rise to the identification of a crossover frequency $\omega_\dagger\!\sim\!L^{-\frac{2}{\dbar+2}}$ in a system of linear size $L$ in $\dbar\!>\!2$ dimensions, above which the notion of discrete phonon bands becomes ill-defined. Instead, above $\omega_\dagger$ phonons are expected to cover the entire frequency axis, leaving no holes or gaps. A first implication of these results is that in the frequency range $\omega\!<\!\omega_\dagger$, where phonon bands are well-defined, phonons exhibit a finite scattering lifetime that scales with $(\Delta\omega)^{-1}$, as we directly demonstrate by dynamic calculations for harmonic disordered lattices.

We further discussed a key implication of our results on the persistence of harmonic quasilocalized glassy modes (QLGMs) --- recently shown to populate the low-frequency tails of the spectrum of 3D structural glasses --- in the thermodynamic limit. We presented extensive computer simulation data that suggest that QLGMs loose their quasilocalized nature if they occur within frequency intervals occupied by phonon bands, due to hybridizations and mixing with those phonons. This, in turn, implies that a coexistence frequency window of phonons and harmonic QLGMs opens at intermediate system sizes, explaining the observations of~\cite{ikeda_pnas,Ikeda_PRE_2018} that report coexistence of these two types of low-frequency excitations, distinguished by their participation ratio. Our results further indicate that the said coexistence frequency window vanishes in the thermodynamic limit, seriously questioning the claim in~\cite{ikeda_pnas,Ikeda_PRE_2018} that coexistence persists in the continuum (i.e.~thermodynamic) limit.

Our work opens up various directions for future investigations. For example, the prefactor in the main theoretical result in Eq.~\eqref{prediction}
\begin{equation}
\chi \equiv \frac{\Delta \omega \sqrt{N}}{\omega\sqrt{n_q}} \,,
\end{equation}
might offer a general, dimensionless quantifier of disorder in any condensed matter system featuring Goldstone modes. In the structural glasses analyzed in this work we found $\chi\!\sim\!{\cal O}(1)$ (see Fig.~\ref{glass}), whereas in the positionally-disordered (glass-derived) Hookean-spring networks we found $\chi\!\sim\!{\cal O}(10^{-1})$ (see Fig.~\ref{disordered_networks}). Future investigations should resolve the relative variations observable in $\chi$ under different preparation protocols in structural glasses and disorder realizations in disordered lattices/crystals. Moreover, the prediction that the phonon damping coefficient should cross over  at $\omega_\dagger$ from being proportional to $\Delta\omega$ to following the Rayleigh scattering scaling (or the modified-Rayleigh scattering scaling in glasses~\cite{Gelin2016}) should be systematically tested.

Several recent efforts, e.g.~\cite{Korkusinski_2007,Popescu_2012,Andrew_2017,Sluiter_high_entropy_alloys_2017,PRB_band_unfolding_method_2017}, are devoted to resolving disorder-induced variations in the statistics of phonons of a prescribed wavevector. These studies exclusively focus on frequencies $\omega\!\gg\!\omega_\dagger$, and show that the major effect of disorder is observed at intermediate to high frequencies. While our theoretical approach is not valid above the crossover frequency, where the notion of discrete phonon bands becomes ill-defined, it would be interesting to resolve whether, if at all, our findings bare relevance to this closely related question.

In our theoretical treatment of the phonon band broadening we assumed that the disorder in the system is not spatially correlated; it would be interesting to understand the effect of spatial correlations of structural disorder on our results. In addition, other applications to a broad range of phenomena involving phonons in disordered systems should be systematically explored.

\acknowledgements
We warmly thank Gustavo D\"uring and Moshe Schechter for fruitful discussions. E.~L.~acknowledges support from the Netherlands Organisation for Scientific Research (NWO) (Vidi grant no.~680-47-554/3259). E.~B.~acknowledges support from the Minerva Foundation with funding from the Federal German Ministry for Education and Research, the William Z.~and Eda Bess Novick Young Scientist Fund and the Harold Perlman Family.

\begin{appendices}

\section{Degenerate perturbation theory}
\label{perturbation_theory_appendix}

In this Appendix we briefly derive Eqs.~\eqref{eq:degenerate_perturbation_theory}-\eqref{eq:degenerate_perturbation_theory_1}, which emerge from standard degenerate perturbation theory. The latter formalism can be found in many Quantum Mechanics textbooks~\cite{footnote3} and is briefly repeated here for completeness. We start with $n_q$ degenerate eigenvectors $\ket{\B \psi_i}$, with $i\!=\!1,2,\ldots,n_q$, corresponding to an eigenvalue $\omega^2$, and denote all other eigenvectors by $\ket{\B \psi_k}$, with $k\!=\! n_q+1,n_q+2,\ldots,\dbar{N}$. The unperturbed Hessian matrix is denoted by ${\calBold M}^{(0)}$ such that any linear combination $\ket{\B \Psi}\!\equiv\!\sum_{m=1}^{n_q} \beta_m \ket{\B \psi_m}$ is an eigenvector
\begin{equation}
\label{eqSM:degenerate_vectors}
{\calBold M}^{(0)}\ket{\B \Psi} = \omega^2 \ket{\B \Psi} \ ,
\end{equation}
where the relevant combination $\ket{\B \Psi}$ for the perturbation theory is not known a priori.

The presence of disorder modifies the Hessian matrix, which now reads ${\calBold M}\={\calBold M}^{(0)}+\delta{\calBold M}$, where the disorder-induced perturbation $\delta{\calBold M}$ gives rise to corrections $\delta\omega^2$ to the eigenvalue $\omega^2$ and $\ket{\delta \B \Psi}$ to the (still unknown) eigenvector $\ket{\B \Psi}$. Perturbing Eq.~\eqref{eqSM:degenerate_vectors} accordingly and keeping terms to linear order, we obtain
\begin{equation}
\label{eqSM:perturbed_eigenproblem}
{\calBold M}^{(0)}\ket{\delta \B \Psi} + \delta{\calBold M}\ket{\B \Psi} = \omega^2\ket{\delta \B \Psi} + \delta\omega^2 \ket{\B \Psi} \ .
\end{equation}
The correction to the eigenvector $\ket{\delta \B \Psi}$ can be expressed in terms of the eigenvectors {\em outside} the degenerate band
\begin{equation}
\label{eqSM:eigenvector_correction}
\ket{\delta \B \Psi} = \sum_{l=n_q+1}^{\dbar{N}} \alpha_l \ket{\B \psi_l} \ .
\end{equation}
Substituting Eq.~\eqref{eqSM:eigenvector_correction} in Eq.~\eqref{eqSM:perturbed_eigenproblem} and contracting with an eigenvector $\bra{\B \psi_i}$ {\em within} the degenerate band, we obtain
\begin{equation}
\label{eqSM:degenerate_perturbation_theory_result}
\sum_{j=1}^{n_q}\bra{\B \psi_i}\delta{\calBold M}\ket{\B \psi_j} \beta_j = \delta\omega^2 \beta_i  \ ,
\end{equation}
which is identical to Eqs.~\eqref{eq:degenerate_perturbation_theory}-\eqref{eq:degenerate_perturbation_theory_1}. Note that if instead we contract with an eigenvector $\bra{\B \psi_l}$ {\em outside} the degenerate band, we can calculate the coefficients $\alpha_l$ and hence the correction to the eigenvector $\ket{\delta \B \Psi}$, though we do not study the latter in this paper.

\section{Numerical methods}
\label{appendix}
In this Appendix we describe the procedures we have followed in the analysis of phonon band widths, we describe how the positionally-disordered networks were created, and we describe the model glass-former we employed.

\subsection{Phonon band widths calculation}
\begin{figure}[!ht]
\centering
\includegraphics[width = 1.0\textwidth]{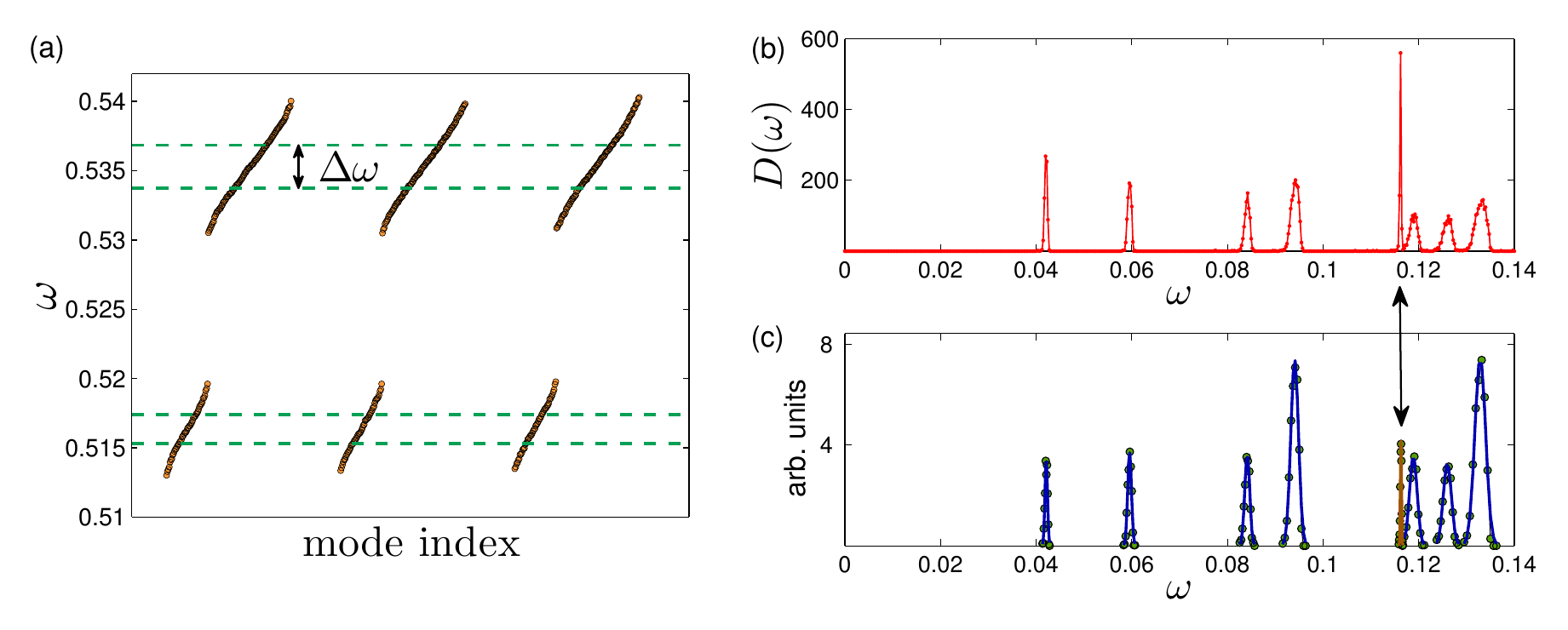}
\caption{\footnotesize In all of the systems of relaxed Hookean springs studied here, the phonon bands can be cleanly identified at low enough frequencies. In panel (a) we show two subsequent bands calculated in three instances of our 3D disordered Hookean spring networks with $N\!=\!64000$. The band widths $\Delta\omega$ are obtained by calculating the standard deviation (over disorder realizations) of the members of each band. (b),(c) Demonstration of how phonon band widths are extracted from the spectra of our computer glasses. We show data for the first few phonon bands calculated for glasses of $N\!=\!409600$ particles in 2D. Panel (b) shows the raw histograms of the eigenfrequencies, collected over several independently-quenched glassy samples. Panel (c) displays the calculated Gaussian fits; notice that for these fits we evaluate the histogram for each peak separately, using the same number (12) of bins for each peak. This explains their different heights compared to the peaks of the raw histogram of panel (b). The peak marked by the arrow at $\omega\!\approx\!0.116$ represents the first sound modes, which explains its minute width.}
\label{getting_widths}
\end{figure}

The calculation of vibrational modes was carried out using the linear algebra package ARPACK \cite{arpack}. In the majority of the calculations performed, extracting phonon widths from the spectra can be done by directly identifying the bands and calculating the standard deviation of the frequencies associated with the members of each band, averaged over several ($100$ or $200$) realizations of the disorder. This is possible since the bands at low frequencies are well-separated, as demonstrated in Fig.~\ref{getting_widths}a.

In our 2D and 3D model glasses a sharp identification of bands is not possible due to the presence of non-phononic modes at low frequencies. We therefore calculated the histogram of the vibrational frequencies, and fitted a Gaussian curve to each peak that was cleanly distinguishable, as demonstrated in Figs.~\ref{getting_widths}b,c. For our 3D glasses, we followed the approach of \cite{ikeda_pnas} and calculated histograms of vibrational frequencies after filtering phonons from non-phononic modes by discarding of modes with participation ratios $e\!<\!0.03$. The standard deviation of the Gaussian fits are reported as the phonon band widths.

\subsection{Disordered networks of Hookean springs}
To create disordered networks of Hookean springs, we utilized the computer glass model that is described in the next subsection. We set the center of each particle as a node, and connected a relaxed Hookean spring of unit stiffness between every pair of interacting particles of the original glass. The mean connectivities of the resulting spring networks we obtained following this procedure are $z\!\approx\!6.5$ in 2D, and $z\!\approx\!16.5$ in 3D, i.e.~very far from the Maxwell threshold 2$\dbar$, and, subsequently, far from the unjamming point~\cite{ohern2003}.

\subsection{Model glass-former}
We employed a 50:50 binary mixture of `large' and `small' particles that interact via a purely-repulsive inverse-power-law ($\sim\!r^{-10}$) pairwise interaction potential. Details of this model system can be found in~\cite{protocol_prerc}. In terms of the microscopic units of energy $\epsilon$ and time $\tau$, the system undergoes a computer glass transition at a temperature $T\!\approx\!0.50\epsilon/k_B$. Systems were initially equilibrated in the high temperature liquid phase at $T\!=\!1.0\epsilon/k_B$, before quenching them to zero temperature at a rate of $\dot{T}\!=\!10^{-3}\epsilon/(k_B\tau)$.

\section{QLGMs within phonon bands}
\label{nonlinear_modes_appendix}

In this Appendix we first explain how the estimation of the number of QLGMs we would expect to find in a frequency interval $\omega_1\!\le\!\omega\!\le\!\omega_2$ is made. Recall that frequencies associated with QLGMs are distributed according to a universal $\omega^4$ law. The prefactor of the $\omega^4$ law, $c_g$, can be extracted from the data of Fig.~\ref{hybridization_fig} by e.g.~counting the total number $n_{\mbox{\tiny QLGM}}$ of QLGMs that were measured over $n_{\mbox{\tiny samples}}$ glassy samples between the two frequencies $\omega_1$ and $\omega_2$. That is, we have
\begin{equation}\label{prefactor}
c_g \approx \frac{5 n_{\mbox{\tiny QLGM}}}{n_{\mbox{\tiny samples}}\dbar{N}(\omega_2^5 -  \omega_1^5)}\,.
\end{equation}
We find $c_g\!\approx\!7.6\!\times\!10^{-4}$ by considering the number of hits between $\omega_1\=0.27$ and $\omega_2\=0.33$, see Fig.~\ref{hybridization_fig}. The number of QLGMs expected to fall between the dashed horizontal lines of Figs.~\ref{hybridization_fig}b,c is then easily obtained by inverting Eq.~(\ref{prefactor}) in favor of $n_{\mbox{\tiny QLGM}}$, using the previously extracted prefactor $c_g$.  We followed this procedure to reach the expectation that we would observe $\approx\!3.3$ QLGMs within the first phonon band, and $\approx\!27$ QLGMs within the second phonon band.

\begin{figure}[!ht]
\includegraphics[width = 0.75\textwidth]{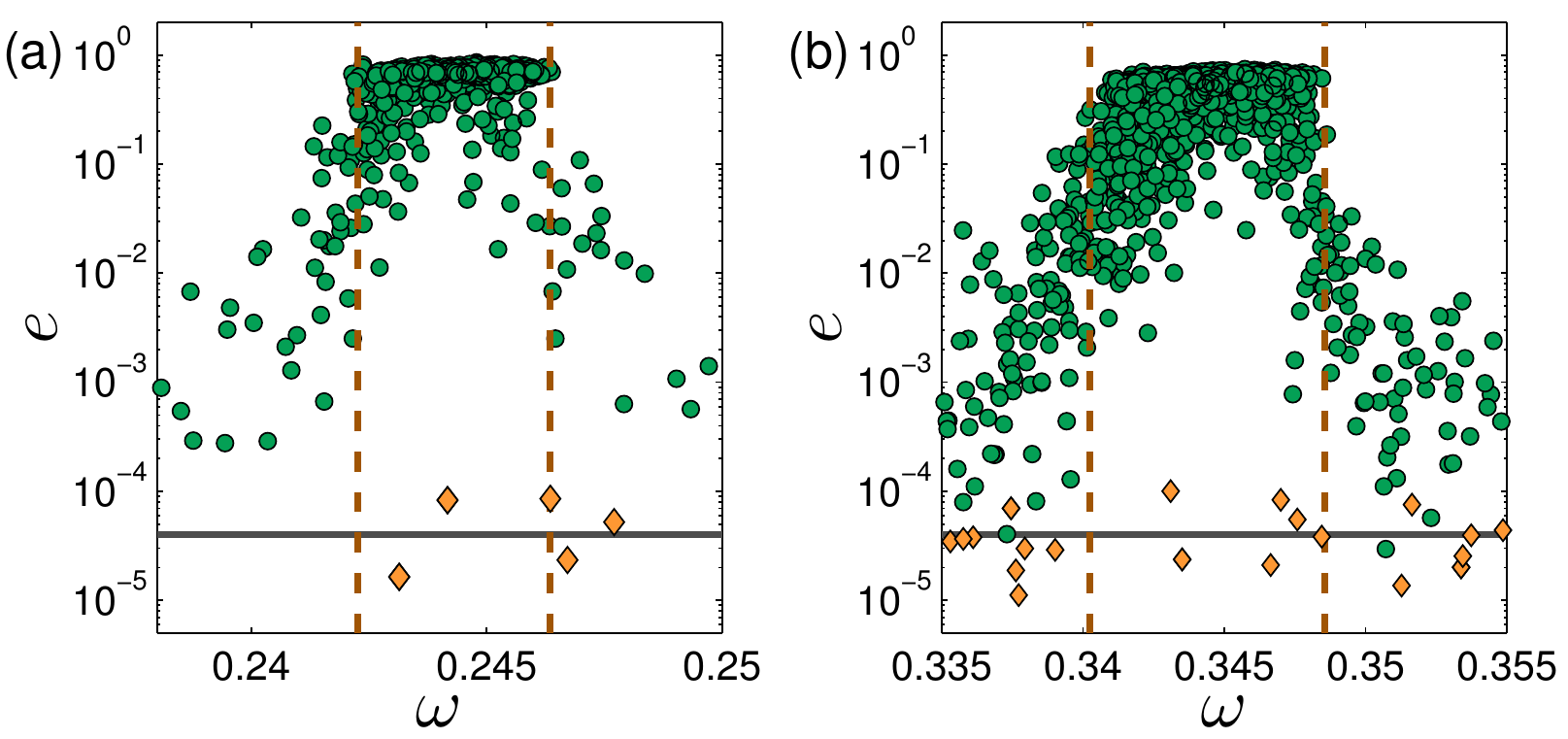}
\caption{\footnotesize The green circles mark the participation ratio $e$ of vibrational modes (see definition in Eq.~(\ref{participation_ratio_definition}) in the main text), plotted against their frequency $\omega$. Data were obtained in systems of $N\!=\!10^6$ particles in 3D; the left and right panels show data for the first and second phonon bands, as seen in Fig.~\ref{hybridization_fig}. The orange diamonds represent nonlinear QLGMs (see text for definitions and details), whose participation ratio precisely matches that of QLGMs outside of phonon bands, as indicated by the horizontal continuous lines, which are deduced from Fig.~\ref{hybridization_fig}a. This data set demonstrates that the soft, quasilocalized excitations with frequencies that fall inside the intervals occupied by phonons are still embedded in the microstructure, despite that they cannot assume the form of harmonic vibrational modes.}
\label{nonlinear_modes}
\end{figure}

We also calculated nonlinear QLGMs; following the definitions put forward in~\cite{SciPost2016}, nonlinear QLGMs $\mathBold{\pi}$ are solutions to the equation
\begin{equation}\label{eq:quartic_modes}
{\calBold M}\cdot\mathBold{\pi} = \frac{{\calBold M}:\mathBold{\pi}\mathBold{\pi}}{\mathBold{{\cal U}''''}::\mathBold{\pi}\mathBold{\pi}\mathBold{\pi}\mathBold{\pi}}\mathBold{{\cal U}''''}\tripleCdot\mathBold{\pi}\mathBold{\pi}\mathBold{\pi}\,,
\end{equation}
where the symbols $:,\tripleCdot$ and $::$ denote double, triple and quadruple contrations, respectively, ${\calBold M}\!\equiv\!\frac{\partial^2{\cal U}}{\partial\xv\partial\xv}$, and $\mathBold{{\cal U}''''}\!\equiv\!\frac{\partial^4{\cal U}}{\partial\xv\partial\xv\partial\xv\partial\xv}$ are the second and fourth order tensors of spatial derivative of the potential energy ${\cal U}$, respectively. For each member of the ensemble of glassy samples of $N\!=\!10^6$ particles in 3D we calculated a single solution to Eq.~(\ref{eq:quartic_modes}), following the methods explained in \cite{plastic_modes_prerc}. The frequency of each solution $\mathBold{\pi}$ is defined as $\omega\!\equiv\!\sqrt{{\calBold M}:\mathBold{\pi}\mathBold{\pi}}$, in analogy with the frequencies of vibrational modes.

Nonlinear QLGMs were shown in \cite{SciPost2016} to (i) possess very small frequencies, (ii) to be quasilocalized in the same fashion as harmonic QLGMs, and (iii) to be entirely indifferent to the presence of phonons of similar frequencies. For these reasons, they serve as a useful tool to identify QLGMs within the frequency intervals occupied by phonon bands. In Fig.~\ref{nonlinear_modes} we plot the same data as in Fig.~\ref{hybridization_fig}b,c, but this time superimpose the frequencies of nonlinear QLGMs (marked by orange diamonds in the figure) detected in the same glassy samples. Since methods to exhaustively detect all nonlinear QLGMs are still unavailable, the number of nonlinear QLGMs we detected within the frequency intervals occupied by the two first phonon bands falls below our prediction as spelled out above. Notwithstanding, the demonstrated existence of nonlinear QLGMs within these frequency intervals unequivocally shows that (i) QLGMs do not assume the form of harmonic vibrations within phonon band frequency intervals due to hybridizations, and (ii) soft quasilocalised excitations whose frequencies fall within those intervals are certainly embedded in the microstructure.

\section{The crossover frequency $\omega_\dagger$ in 2D}
\label{sec:2D}

\begin{figure}[!ht]
\includegraphics[width = 0.7\textwidth]{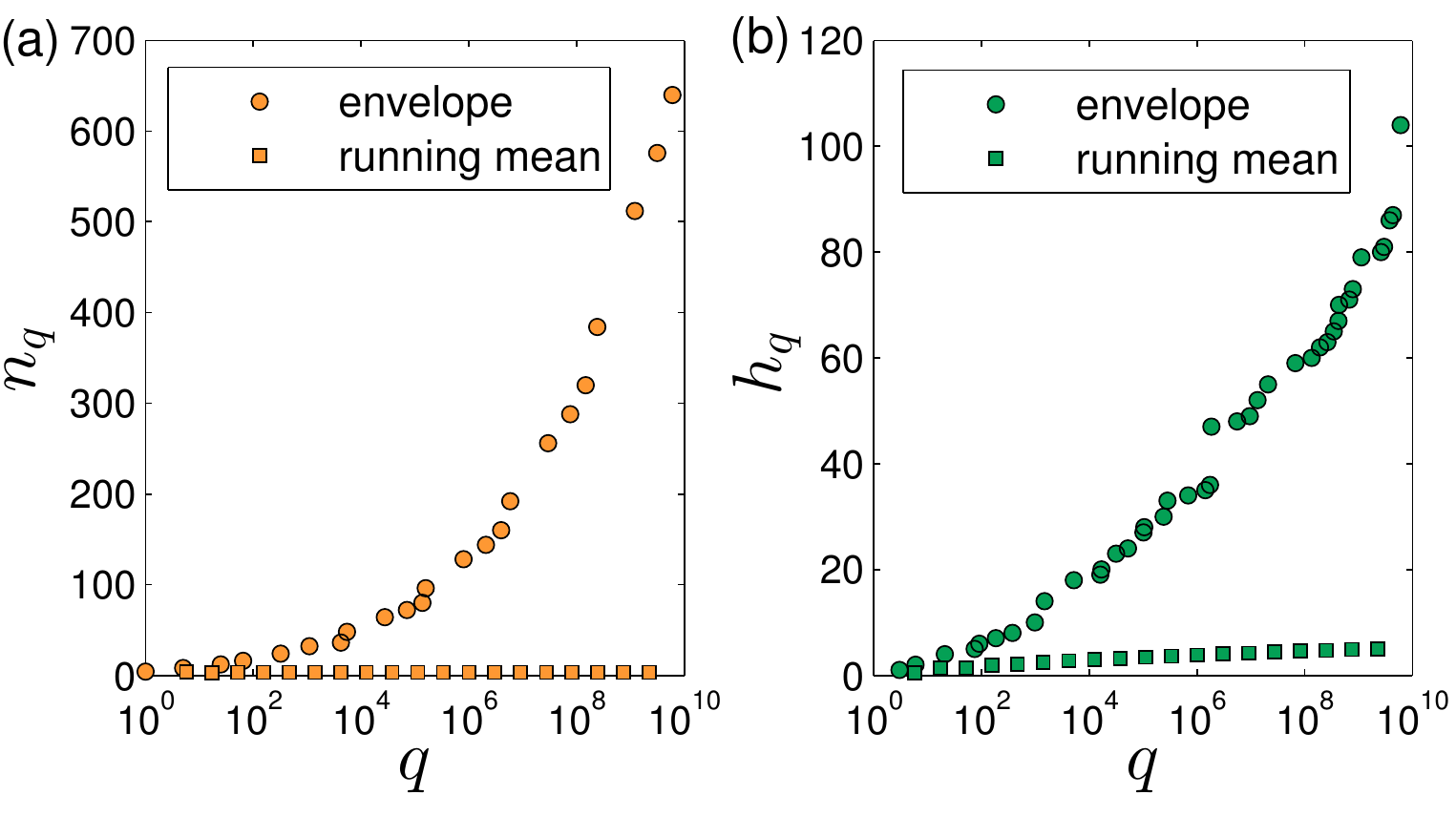}
\caption{\footnotesize (a) Number of solutions $n_q$ to the integer sum of squares problem as given by Eq.~(\ref{eq:sum_of_squares}), for $\dbar\!=\!2$. The circles represent the envelopes defined as pairs $(q,n_q)$ such that $n_q\!>\!n_{q'}$ for all $q'<q$, while the squares represent running averages binned over $q$. (b) Intervals $h_q$ of consecutive integers $q'$ for which $n_{q'}\!=\!0$, defined for integers $q$ that feature $n_q\!>\!0$, see text for further details. The envelopes and running averages are as defined in panel (a).}
\label{sum_of_integer_squares}
\end{figure}

In Sect.~\ref{sec:omega_dagger} we derived the scaling of the crossover frequency $\omega_\dagger\!\sim\!L^{-2/(\dbar+2)}$ for a disordered solid of linear size $L$ in $\dbar\!>\!2$ spatial dimensions, above which phonon bands overlap and merge, leaving the density of vibrational modes free of gaps. Resolving the $L$-dependence of $\omega_\dagger$ in 2D is less straightforward, as a result of the erratic behavior of the degeneracy $n_q$ of the $q\th$ phonon band, which hinders a scaling estimation of the broadening of phonon band widths $\Delta\omega$. In particular, despite that
\begin{equation}
\sum_{q'=0}^q n_{q'} \sim q\,,
\end{equation}
$n_q$ is not a constant function of $q$ in 2D. Instead, $n_q$ features two trends with increasing $q$, illustrated in Fig.~\ref{sum_of_integer_squares}; first, the envelope of $n_q$ (see figure caption for definition) increases with increasing $q$, as shown in the left panel of Fig.~\ref{sum_of_integer_squares}. On the other hand, the number of integers $q$ for which $n_q\!=\!0$ grows with increasing $q$ as well~\cite{landau_1908}, many of which occupy consecutive intervals of integers, referred to here as \emph{holes}. The latter, denoted by $h_q$, are defined for integers $q$, that feature finite degeneracies $n_q\!>\!0$, as the number of \emph{consecutive} integers $q'\!>\!q$ for which $n_{q'}\!=\!0$. In Fig.~\ref{sum_of_integer_squares}b we plot the envelope of $h_q$, which is seen to grow in a similar manner as the envelope of $n_q$. The appearance of growing holes in $n_q$ hinders the clean estimation of the scaling of the frequency gaps between consecutive phonon bands with band index $q$ and system size $L$, as we have done for $\dbar\!>\!2$ in Eq.~\eqref{eq:gaps} of Sect.~\ref{sec:omega_dagger}.

In Fig.~\ref{sum_of_integer_squares} we have also plotted the running averages of the degeneracies $n_q$ (squares, left panel) and of the holes $h_q$ (squares, right panel), binned over logarithmic intervals of the band index $q$. The running average of $n_q$ is a constant of order one, whereas the running average of $h_q$ grows slower than logarithmically. We therefore neglect their weak $q$-dependence altogether, and assume that $n_q$ and $h_q$ are constants.

Under the assumptions discussed above, which we cannot fully justify, our estimation of the broadening of phonon band follows verbatim Eqs.~(\ref{eq:foo00})-(\ref{eq:foo01}), and thus for 2D disorder solids we expect the crossover frequency $\omega_\dagger\!\sim\!L^{-1/2}$. This prediction is in fact consistent with Eq.~(\ref{eq:foo01}), when $\dbar\!=\!2$ is used.

\vspace{0.1cm}

\end{appendices}

%

\end{document}